\documentclass[prc,aps,twocolumn,preprintnumbers,superscriptaddress]{revtex4}

\usepackage[normalem]{ulem}
\usepackage{mathrsfs}
\usepackage{amssymb}
\usepackage{graphicx}
\usepackage{dcolumn}
\usepackage{bm}
\usepackage{graphicx}
\usepackage{float}
\usepackage{longtable}
\usepackage{amsmath}
\usepackage{color}
\usepackage{diagbox}
\usepackage{multirow}
\usepackage{braket}
\usepackage{booktabs}
\usepackage[colorlinks=true,linkcolor=blue,filecolor=blue, urlcolor=blue,citecolor=blue]{hyperref}
\usepackage{threeparttable}
\usepackage{amsmath}

\newfont{\largemi}{cmmi10}
\newfont{\smallmi}{cmmi6}

\draft

\def\eqref#1{Eq.~(\ref{#1})}
\def\figref#1{Fig.~\ref{#1}}

\fontsize{20}{20}

\begin{document}
	

\title{Shell model description of the \(N=82\) isotonic chain with a new effective interaction}
    
	\author{Y. X. Yu}
	\affiliation{School of Physics Science and Engineering, Tongji University, Shanghai 200092, China}
	
	\author{Q. Y. Chen}
	\affiliation{School of Physics Science and Engineering, Tongji University, Shanghai 200092, China} 
	\author{Chong Qi}
	\affiliation{Department of Physics, KTH Royal Institute of Technology, 10691 Stockholm, Sweden} 
	    \affiliation{ Academy of Romanian Scientists, Ilfov Street, no.3, 050044, Bucharest, Romania}%
	\author{G. J. Fu}
	\email{gjfu@tongji.edu.cn}
	\affiliation{School of Physics Science and Engineering, Tongji University, Shanghai 200092, China}

	\date{\today}

\begin{abstract}

In this work, we present a systematic study of low-lying states and electromagnetic properties of the semi-magic $N = 82$ isotonic chain with proton number $Z=51$-77, using the full configuration interaction shell model with a newly developed high-quality effective interaction. 
The calculations are performed in a large model space that includes all proton orbitals between $Z = 50$ and 82: $0g_{7/2}$, $1d_{5/2}$, $1d_{3/2}$, $2s_{1/2}$, and $0h_{11/2}$. 
The effective interaction
is derived through the principal component analysis approach, starting from 160 two-body matrix elements and 5 single-particle energies and considering up to 30 degrees of freedom. Those are optimized by fitting to 204 available experimental energy levels. 
The resulting root-mean-square deviation is as low as 102 keV. 
The new interaction successfully reproduces the binding energies, low-lying spectra, electric quadrupole transition probabilities $B(E2)$, and magnetic dipole moments across both even-even and odd-mass isotones.
The nuclear structure of low-lying states is analyzed in detail.
Additionally, predictions are made for several more proton-rich nuclei beyond current experimental reach, including $^{155}\mathrm{Ta}$, $^{156}\mathrm{W}$, $^{157}\mathrm{Re}$, $^{158}\mathrm{Os}$, and $^{159}\mathrm{Ir}$.

\end{abstract}
	
	\vspace{0.4in}
	
	\maketitle

\section{INTRODUCTION}

The full-configuration-interaction shell model, more commonly known as the nuclear shell model, is one of the most successful and versatile microscopic approaches for studying atomic nuclei. 
Shell model calculations within a defined model space, using either realistic interactions (derived, e.g., from many-body perturbation approaches, or through the valence-space in-medium similarity renormalization group method starting from realistic nuclear forces) or phenomenological interactions [e.g., optimized two-body matrix elements (TBMEs) or monopole terms], have yielded reliable and predictive results~\cite{talmibook,sm-talmi,BROWN2001517,RevModPhys.77.427,S0218301323300035}.
In the phenomenological case, excellent agreement with experiment can often be achieved with minimal modifications to realistic interactions,    
offering a practical tool for both theoretical studies and experimental interpretation.
For nuclei in the light- and medium-mass regions, shell model calculations have proven effective in explaining the properties of low-lying states. For instance, the empirical Hamiltonians USDA and USDB accurately reproduce low-lying energy levels, binding energies, and electromagnetic properties for nuclei in the $sd$ shell \cite{PhysRevC.74.034315}. Similarly, the GXPF1 \cite{PhysRevC.65.061301} and KB3G \cite{POVES2001157} interactions have been highly successful in describing a wide range of properties in $pf$-shell nuclei. The JUN45 interaction, derived from the realistic Bonn-C nuclear force and empirically optimized to match experimental data, has been widely used in $f_{5}pg_{9}$-shell calculations and effectively reproduces excitation energies and magnetic moments of nuclei up to $N = Z = 50$, particularly those involving the $g_{9/2}$ orbital \cite{PhysRevC.80.064323}.

There has been significant recent experimental and theoretical interest in the structure of nuclei with proton or neutron numbers between 50 and 82, corresponding to the $g_7dsh_{11}$ shell~\cite{PhysRevC.109.054317,cederlof2023lifetime,
STORBACKA2024138822,
PhysRevC.96.051304,
PhysRevC.94.024321,
PhysRevC.91.061304,
PhysRevC.88.044332,
PhysRevC.87.031306,
PhysRevC.87.014308,
PhysRevC.86.044323,
Qi_2016,FAESTERMANN1979190,Bhoy_2022}. 
Particular attention has been devoted to the extended chains of Sn and Te isotopes, as well as the $N = 80$ and 82 isotonic chains.
These studies show the robustness of the $Z = 50$ and $N = 82$ shell closures, and the subshell effects observed at $Z = 64$ (or $N=64$), which arises from the energy gap separating the $0g_{7/2}, 1d_{5/2}$ orbitals from the $1d_{3/2}, 2s_{1/2}, 0h_{11/2}$ orbitals.
Increasing experimental and theoretical attention has also been directed toward proton-rich nuclei near or beyond the drip line, such as $^{155}\mathrm{Ta}$ and $^{150}\mathrm{Yb}$. 
Reliable theoretical predictions are essential for understanding nuclear structure in this region and provide valuable guidance for future experimental studies.
Despite the exponential growth in the dimension of many-body bases, shell model calculations remain computationally feasible for intermediate-mass and heavy nuclei near major neutron or proton shell closures.

Several effective interactions have been developed for the $N = 82$ isotonic chain, where the valence protons are active in the $g_7dsh_{11}$ shell.
Notable examples include the JJ56PNA \cite{BROWN2014115} and KH5082 \cite{kuo1971us,PhysRevC.43.602} interactions. 
The KH5082 interaction, developed by Kuo, Herling $et$ $al.$, starts from a free nucleon-nucleon potential and incorporates renormalizations to account for the contributions from outside model space. The CW5082 interaction \cite{PhysRevC.45.1720} was subsequently derived from KH5082 by replacing the proton-proton TBMEs with those from the effective interaction proposed by Kruse and Wildenthal \cite{Kruse}.
The region from $Z = 52$ to 63 has been explored using interactions based on meson-exchange potential models \cite{HOLT1997107,SUHONEN199841}. Previous work has also investigated low-spin excitation energies in the $N = 82$ isotonic chain using time-dependent degenerate linked-diagram perturbation theory \cite{PhysRevC.80.044320}.

In this work, we present a detailed and systematic investigation of the low-lying states in the $N = 82$ isotones through an optimization of the effective Hamiltonian using the principal component analysis (PCA) approach~\cite{PCA1,PCA2,PCA3}. 
A preliminary version of the monopole and multipole optimized effective interaction was introduced in Ref.~\cite{CQY}, which reasonably reproduces the systematic trend of the $2^+$ excitation energies.
That interaction was constructed starting from the realistic JJ56PNA interaction~\cite{BROWN2014115}, with all proton-proton TBMEs scaled by a factor of 0.82, and further optimized by adding correction terms to the monopole interaction, the multipole interactions of ranks 2, 3, and 5, as well as the octupole-pairing interaction.
In the present work, we achieve excellent agreement with experimental data and resolve previous systematic discrepancies, including consistent overestimations of the $3^-$ excitation energies and deviations in the predicted ground-state spins and parities of the heavier odd-$A$ isotones.
We systematically investigate binding energies, low-lying spectra, electric quadrupole reduced transition probabilities $B(E2)$, and magnetic dipole moments $\mu$. 
We further analyze the dominant configurations that characterize the structure of these low-lying states.

The paper is organized as follows. Sec. II introduces the model space and the PCA-based optimization of the effective Hamiltonian. 
Sec. III presents the calculated results, compares them with available experimental data, and discusses the underlying structures of the low-lying states.
Finally, Sec. IV briefly summaries the results and provides an outlook.

\section{Theoretical framework}

To describe the $N = 82$ isotones, we consider valence protons in the $g_7dsh_{11}$ shell, which includes the $0g_{7/2}$, $1d_{5/2}$, $1d_{3/2}$, $2s_{1/2}$, and $0h_{11/2}$ orbitals, above the doubly magic nucleus $^{132}$Sn taken as the inert core. 
The low-lying states are calculated using the shell model code BIGSTICK \cite{JOHNSON20132761,johnson2018bigstickflexibleconfigurationinteractionshellmodel}. 
The largest $M$-scheme dimension encountered in this work is for $^{148}$Dy, reaching $1.6 \times 10^7$ for even-parity states.

The shell model Hamiltonian is given by
	\begin{equation}
		\begin{aligned} 
			\hat{H}=& \sum_a \varepsilon_a \hat{n}_a+\sum_{J T}\sum_{a \leq b, c \leq d} V_{J T}(a b c d)\\
			& \times \sum_{M T_z} \hat{A}^{(JT)\dagger}_{ M T_z}(a b) \hat{A}^{(JT)}_{ M T_z}(c d) \\
           =& \sum_a \varepsilon_a \hat{n}_a + \sum_{J T} \sqrt{(2J+1)(2T+1)}  \\
			& \times \sum_{a \leq b, c \leq d} V_{J T}(a b c d)  \left[  \hat{A}^{(JT)\dagger}(a b) \times \hat{\tilde{A}}^{(JT)}(c d) \right]^{(0)} , 
		\end{aligned} \label{H}
	\end{equation}
where \(\hat{n}_a\) is the single-particle number operator for the orbit \(a\) with quantum numbers \((n_a, l_a, j_a)\), and 
	\begin{equation}
		\begin{aligned} 
			\hat{A}^{(JT)\dagger}_{ M T_z}(a b) &=\frac{ 1}{\sqrt{1+\delta_{ab}}} (	\hat{a}_a^\dagger \times 	\hat{a}_b^\dagger)^{(JT)}_{ M T_z}
		\end{aligned}
	\end{equation}
is a normalized two-particle creation operator coupled to good spin and isospin. $\hat{A}^{(JT)}_{ M T_z}$ is the annihilation operator, with $\hat{\tilde{A}}^{(JT)}$ its spherical-tensor form.
In this work, the Hamiltonian involves 5 proton single-particle energies (SPEs) $\varepsilon_a$ and 160 proton-proton TBMEs $V_{J T}(a b  c d)$ with $T=1$.
We apply the standard mass scaling factor $(A/134)^{-1/3}$ to the TBMEs, which is frequently adopted in shell model calculations.
The Coulomb interaction was not explicitly included, but its effects were effectively incorporated through the fitting procedure.
For compact notation, the Hamiltonian in \eqref{H} can be rewritten in the following form:
	\begin{equation}
		\hat{H}=\sum_{i=1}^{t} a_i 	\hat{O}_i,
	\end{equation}
where $\hat{O}_i$ represents either the particle number operator $\hat{n}_a$ or the two-body operator $ \hat{A}^{(JT)\dagger}_{ M T_z}(a b) \hat{A}^{(JT)}_{ M T_z}(c d)$, and the coefficients $\vec{a}$ are the corresponding parameters, i.e., $\varepsilon_a$ and $ V_{J T}(a b c d)$.

We optimize the parameters $\vec{a}$ using the least square fitting, combined with a PCA method~\cite{PCA1,PCA2,PCA3} to truncate the parameter space. 
We use $N$ to denote the number of data points to be fitted, and $t$ the total number of input parameters.
The experimental value of the $k$-th data is denoted by $y^k$, and the corresponding theoretical formula by $f^k=f(k;\vec{a})$. 
At the $s$-th iteration, the parameters are given by $\vec{a}^{(s)}$.
We minimize the weighted least-square loss function:
	\begin{eqnarray}
		\nabla_{\Delta\vec{a}} \sum_{k=1}^N \omega_k (y^k-f^k)^2 = 0, \label{eq4} \\
		f^k \approx f^{k(s)} + \vec{g}^{k(s)} \cdot \Delta\vec{a}, \label{eq5}
	\end{eqnarray}
where the gradient $\vec{g}^{k(s)}$ is defined by $g^{k(s)}_i \equiv  \partial f^{k} / \partial a_i |_{\vec{a}^{(s)}}$,
$\Delta\vec{a}$ represents the change of the parameters,
and $\omega_k$ is the weight associated with the $k$-th data point.
All weights are set to 1, except for the binding energies of \(^{152}\textrm{Yb}\) and \(^{153}\textrm{Lu}\), which are both down-weighted to $0.5$ due to their large experimental uncertainties.
Combining Eqs. (\ref{eq4}) and (\ref{eq5}) yields the linear equations:
	\begin{eqnarray}
		A \Delta \vec{a} = \vec{b}, \label{fit}
	\end{eqnarray}
	with
	\begin{eqnarray}
		A_{ij}=\sum_{k=1}^N  \omega_k g^{k(s)}_i g^{k(s)}_j, \\
		\vec{b} =\sum_{k=1}^N \omega_k (y^k-f^{k(s)})\vec{g}^{k(s)}. 
	\end{eqnarray}

The standard least square fitting would typically stop at this stage. But to implement the PCA, we proceed by diagonalizing the error matrix $D= A^{-1}$:
	\begin{equation}
		{\rm diag}(d_1,d_2, \dots, d_{t})=PDP^{T}, \label{diag}
	\end{equation}
	where the eigenvalues $d_i$ are sorted in ascending order. 
	Using the transformation $P$, \eqref{fit} becomes
	\begin{eqnarray}
		\Delta c_i = d_i e_i,  \label{fit2}
	\end{eqnarray}
where $\Delta \vec{c} \equiv P\Delta \vec{a}$, and $\vec{e} \equiv P \vec{b}$. 
The components $\Delta \vec{c} $ represent the change along the principal direction of the parameter space, referred to as the principal parameters.
These principal parameters are linear combinations of the original parameters $\vec{a}$ and are selected according to their sensitivity in reducing the residuals of the fit.
The most important variations are associated with the components having the smallest uncertainties, i.e., the principal directions corresponding to the smallest eigenvalues of $D$.
The PCA is applied by retaining only the $q$ components associated with the smallest eigenvalues, where $q$ defines the dimension of the truncated parameter space.
This truncation is implemented as follows:
	\begin{equation} \label{pca}
		\Delta c^{\prime}_i=\left\{\begin{array}{cc} 
			d_i e_i ,& \text{if} ~~ i \leqslant q , \\
			0 ,& \text{otherwise},
		\end{array}\right.
	\end{equation} 
where $\Delta c^{\prime}_i$ represents the truncated changes of the principal parameters. 
The components with larger $d_i$ are neglected, ensuring the fit focuses on the most important parameters.
The parameters $\vec{a}$ are then updated using 
	\begin{equation}
		\vec{a}^{(s+1)} = \vec{a}^{(s)} +  P^T \Delta \vec{c}^{\prime}.
	\end{equation}
The above procedures are repeated iteratively until convergence is achieved.
Similar parameter space truncation approaches have been widely used to optimize shell model interactions~\cite{ARIMA196894,RICHTER1991325,HONMA2002134,PhysRevC.65.061301,PhysRevC.70.044314,PhysRevC.74.034315,PhysRevC.80.064323}.
    
In this work, we restrict the parameter space to $q = 30$, {which effectively mitigates potential overfitting issues}. 
Our dataset comprises 204 experimental data points for the $N = 82$ isotones, including 21 nuclear binding energies, 101 excitation energies in even-even nuclei (where available, we focus on the first $2^+, 4^+, 6^+, 8^+, 10^+, 12^+$ states, as well as the $0_2^+, 2_2^+, 4_2^+, 6_2^+, 8_2^+, 10_2^+, 3_1^-, 5_1^-$, and $7_1^-$ states), and 82 excitation energies in odd-$A$ nuclei (for example, in $^{149}$Ho, states include $11/2^-$, $15/2^-$, $19/2^-$, $23/2^-$, $27/2^-$, $1/2^+$, $3/2^+$, $5/2^+$, $7/2^+$, $19/2^+$, and $23/2^+$).
As the starting point, we adopt the initial parameter values from Ref. \cite{CQY}. The optimization, as described above, converges in only three iterations, indicating the stability and efficiency of the fitting procedure. 
The root-mean-square deviation (RMSD) between the experimental values and the shell model results for the 204 data points is 102 keV.

\begin{table}[htbp]
\centering
\caption{Single-particle energies (MeV) relative to the $0g_{7/2}$ orbital. The values given in the parentheses denote the nuclear binding energy difference between $^{133}\textrm{Sb}$ and $^{132}\textrm{Sn}$. }
\label{tab:energy_levels}
\begin{tabular}{|c|c|c|c|}
\hline
Orbital & This work & JJ56PNA \cite{BROWN2014115} & Expt. \\
\hline
$0g_{7/2}$ & 0 $(-9.670)$ & 0 $(-9.667)$  & 0 $(-9.658)$ \\
\hline
$1d_{5/2}$ & 1.117 & 0.962 & 0.962 \\
\hline
$2s_{1/2}$ & 2.728 & 2.340 & $-$ \\
\hline
$1d_{3/2}$ & 2.457 & 2.708 & 2.439 \\
\hline
$0h_{11/2}$ & 2.790 & 2.793 & 2.791 \\
\hline
\end{tabular}
\end{table}

The obtained SPEs are listed in Table \ref{tab:energy_levels}, in comparison with experimental data and those from another effective interaction, JJ56PNA~\cite{BROWN2014115}. 
The values given in parentheses represent the nuclear binding energy difference between $^{133}$Sb and $^{132}$Sn, i.e., the one-proton separation energy.
The calculated SPEs are in good agreement with experimental values. 
The SPE of the $1d_{5/2}$ orbital in this work is 0.16 MeV higher than both the experimental data and the JJ56PNA interaction. 
The SPE of the $2s_{1/2}$ orbital is 0.39 MeV higher than in JJ56PNA; however, no corresponding experimental data are available. 
A relatively large uncertainty is expected due to the limited experimental information available to tightly constrain the SPE of the $2s_{1/2}$ orbital, which has both low occupancy and low degeneracy.

\begin{figure}[h]
		\centering
		\includegraphics[width=1\linewidth]{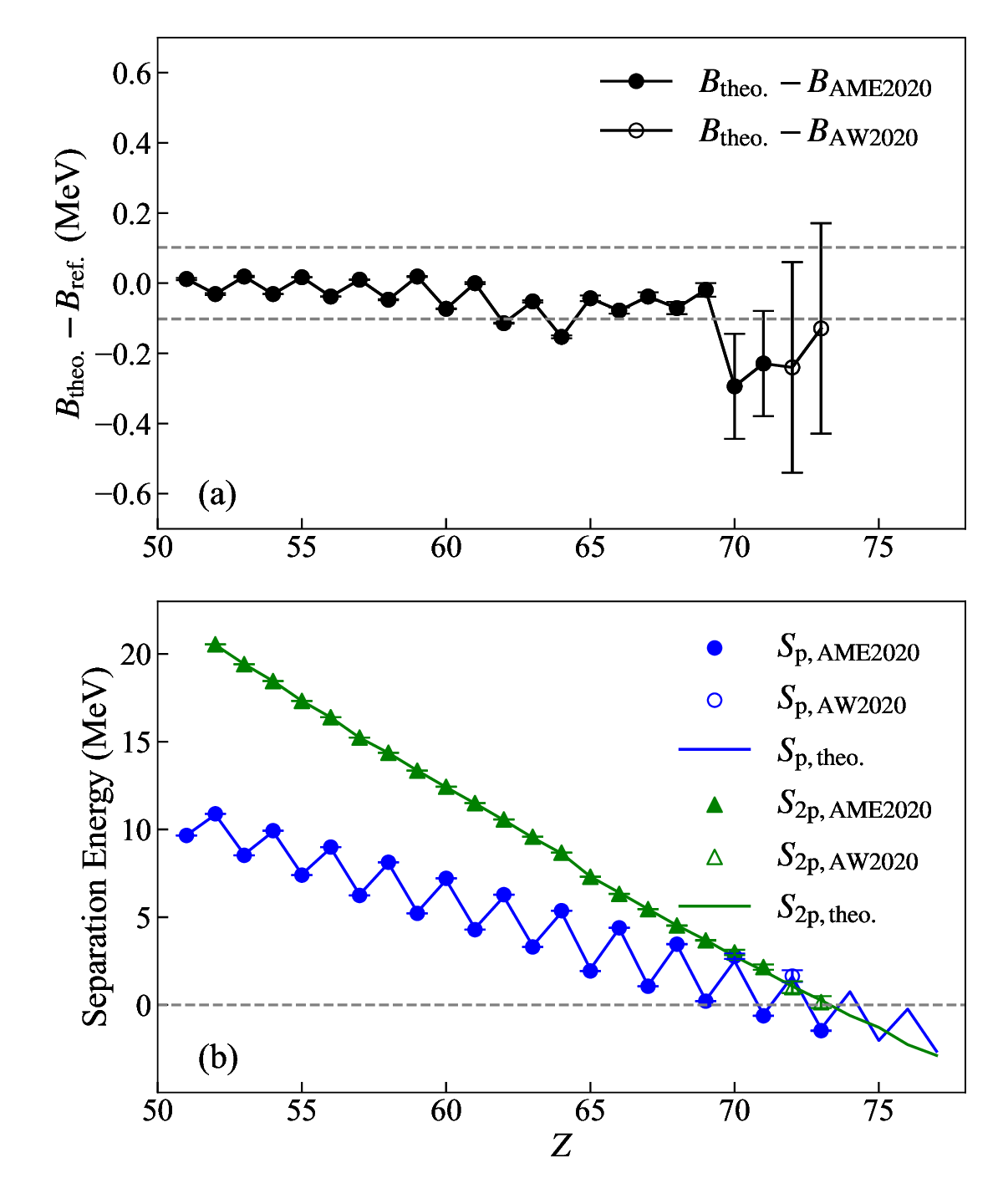}
\caption{
        (a) Difference between the shell model and reference values of the nuclear binding energy ($B_{\rm theo.} - B_{\rm ref.}$) for the $N=82$ isotones.
        The solid dots represents experimental values from the AME2020 database~\cite{Wang_2021}, 
        while the open circles denote the Audi-Wapstra 2020 extrapolation (AW2020)~\cite{Wang_2021}.
        The dashed line in gray indicates the RMSD of 102 keV obtained for all 204 data points included in the fit (see the text).
        (b) One-proton and two-proton separation energies.
        The curves represent the shell model results in this work.        
}
		\label{binding}
\end{figure}

\section{results and discussion}

In this section, we present results for nuclear binding energies, low-lying level spectra, electric quadrupole transition strengths $B(E2)$, and magnetic dipole moments $\mu$ for the even-even and odd-$A$ $N=82$ isotones from $^{134}$Te to $^{159}$Ir.

\subsection{Binding energies and separation energies}

We calculate nuclear binding energies relative to the doubly-magic core $^{132}$Sn. i.e., \(  B(Z,N) - B(^{132}\text{Sn}) \).
Fig.~\ref{binding}(a) compares these binding energies from the AME2020 database~\cite{Wang_2021} with our shell model result. 
For the isotones $^{134}$Te-$^{151}$Tm, the calculation reproduces the experimental values with high accuracy, yielding an RMSD of 59 keV.
The residual deviations exhibit a small odd-even staggering of about 60 keV for nuclei with $Z<70$, indicating that a small part of the odd-even effect is not fully captured.
In comparison, calculations with earlier implementations of shell model interactions~\cite{BROWN2014115,kuo1971us,PhysRevC.43.602,PhysRevC.45.1720}, which were not optimized for binding energies, typically deviate from experiment by about 1 MeV up to several tens of MeV.

For isotones beyond $^{151}$Tm, we report atomic masses (mass excesses), which are directly measured and tabulated experimental quantities, and therefore allow a direct comparison with experiment.
For $^{152}$Yb and $^{153}$Lu, the experimental atomic mass excesses carry large uncertainties, reported as $-46.270(150)$ MeV and $-38.375(150)$ MeV, respectively. These uncertainties may partially account for the relatively larger deviations from the shell model predictions, which are $-45.977$ MeV and $-38.146$ MeV, respectively.
Experimental values for the atomic mass excesses of $^{154}$Hf and $^{155}$Ta are not available, but the 2020 Audi-Wapstra extrapolation (labeled AW2020) ~\cite{Wang_2021} provides estimates of $-32.730(300)$ MeV and $-23.988(300)$ MeV, while the shell model predicts $-32.490$ MeV and $-23.859$ MeV, respectively.
The shell model further provides predictions for the atomic mass excesses of the more proton-rich isotones $^{156}$W-$^{159}$Ir, with values of $-17.335$ MeV, $-8.032$ MeV, $-0.523$ MeV, and $9.393$ MeV, respectively.

We also calculate the one-proton and two-proton separation energies, defined by:
\begin{equation}
\begin{aligned}
   S_{\rm p}(Z, N)=&B(Z, N)-B(Z-1, N), \\
      S_{\rm 2p}(Z, N)=&B(Z, N)-B(Z-2, N) .
\end{aligned}
\end{equation}    
In Fig.~\ref{binding}(b), experimental data for the one-proton and two-proton separation energies are available up to $^{155}$Ta, while the shell model calculation extends to $^{159}$Ir. The one-proton separation energy for $^{153}$Lu is experimentally found to be negative, indicating that this nucleus lies beyond the proton drip line. 
In contrast, both the one-proton and two-proton separation energies for $^{154}$Hf are positive, suggesting that the nucleus remains bound. They are well reproduced by our shell model result.
For $^{155}$Ta, the shell model predicts a negative one-proton separation energy and a positive two-proton separation energy, which points to the possibility of single-proton emission. This behavior has been previously reported in Refs.~\cite{PhysRevC.59.R2975,PhysRevC.75.061302}.
For the recently observed nucleus $^{156}$W, the calculated one-proton separation energy is positive, while the two-proton separation energy is negative ($-0.606$ MeV). 
This indicates the possibility of two-proton emission; however, experimental results show that this decay mode is strongly suppressed, with $\beta^+$ decay being dominant~\cite{PhysRevLett.132.072502,BRISCOE2023138310}.
For the experimentally unmeasured nucleus $^{158}$Os, the shell model predicts a one-proton separation energy of $-0.235$ MeV and a two-proton separation energy of $-2.264$ MeV. This implies that while single-proton emission is hindered, two-proton emission could be a favored decay channel.
In the cases of $^{157}$Re and $^{159}$Ir, the calculated one-proton separation energies are $-2.029$ MeV and $-2.643$ MeV, respectively, while the corresponding two-proton separation energies are $-1.279$ MeV and $-2.878$ MeV.
All these values are negative, indicating that both nuclei are unbound with respect to one- and two-proton emission.

\subsection{Electromagnetic properties}    

\begin{figure}[h]
		\centering
		\includegraphics[width=0.9\linewidth]{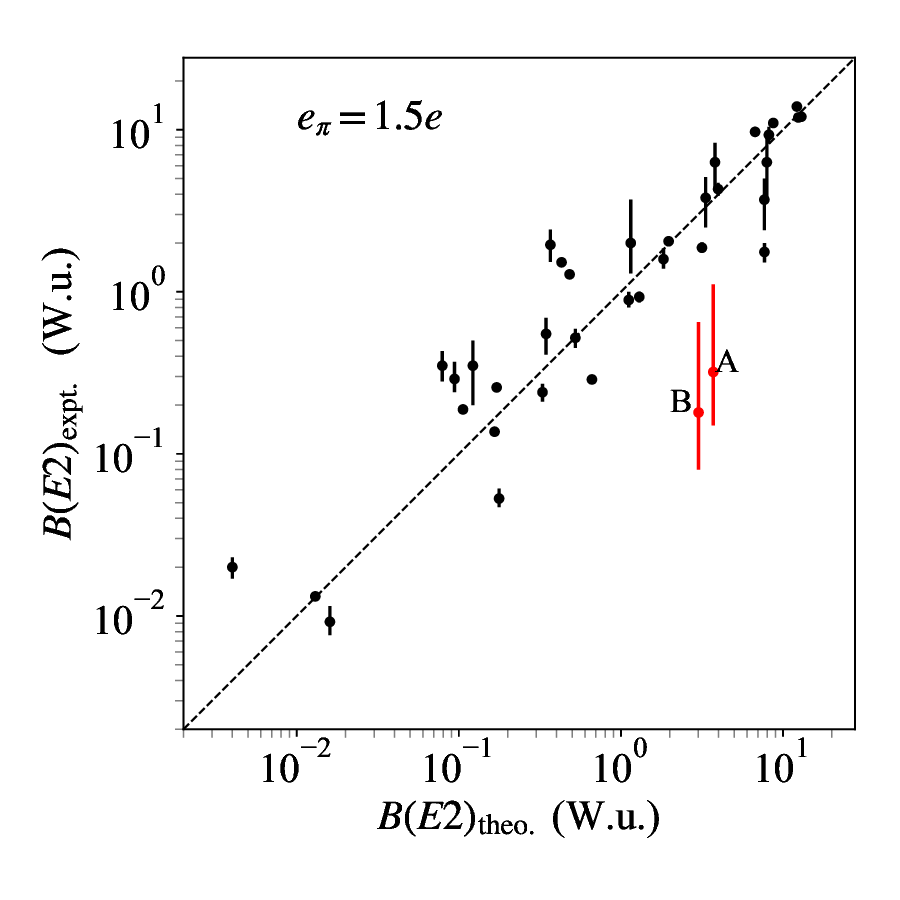}
		\caption{Comparison of \(B(E2)\) values between experiment~\cite{PhysRevC.104.014316,A=135,A=137,A=139,A=141,A=143,A=145,A=147,A=149,A=151,A=153,A=134,A=136,A=138,A=140,A=142,A=144,A=146,A=148,A=150,A=152,A=154} and shell model, shown in a logarithmic scale. 
        The dashed line denotes the line where calculated values equal experimental ones.
        Two transitions show noticeable deviations from this line (marked in red): A. \(8_1^+ \to 6_1^+\) in \(^{138}\textrm{Ba}\) and B. \(8_1^+ \to 6_2^+\) in \(^{138}\textrm{Ba}\). 
        The data point for the \(9/2_1^+ \to 7/2_1^+\) transition in \(^{139}\textrm{La}\) falls outside the limits of this figure.
        }
		\label{E2_log}
\end{figure}

The electric quadrupole reduced transition strength $B(E2)$ is defined by
\begin{equation}
B(E2; J_i \to J_f) = \frac{1}{2J_i + 1} \left| e_{\pi}\langle J_f || \hat{Q}_2 || J_i \rangle \right|^2,
\end{equation}
where $\hat{Q}_2$ is the quadrupole operator,
\begin{equation}
\hat{Q}_{2\mu} =  \hat{r}^2 \hat{Y}_{2\mu},
\end{equation}
with \(e_{\pi}\) denoting the proton effective charge, and $\langle J_f || \hat{Q}_2 || J_i \rangle$ representing the reduced matrix element.


Fig.~\ref{E2_log} compares the calculated \(B(E2)\) values obtained with the shell model using the standard effective charge \(e_{\pi} = 1.5 e\) to 38 experimental values listed in Table \ref{BE2} in Appendix~\ref{appa}. Fig.~\ref{E2_log} provides a global overview of the agreement between theory and experiment, while Table \ref{BE2} gives the numerical values for detailed comparison.
Most calculated points lie along the diagonal in Fig.~\ref{E2_log}, indicating good agreement.
As an example, for \(^{141}\textrm{Pr}\),  the experimental \(B(E2)\) values for the transitions \(9/2_1^+ \to 5/2_1^+\) and  \(9/2_2^+ \to 5/2_1^+\) are \(1.95^{+0.47}_{-0.42}\) W.u. and \(9.3 \pm {1.1}\) W.u., respectively.
The corresponding shell model predictions are 0.367 W.u. and 8.173 W.u.
The calculated excitation energies for the \(9/2_1^+ \) and \(9/2_2^+ \) states are 1.529 MeV and 1.534 MeV, respectively, in good agreement with the experimental values of 1.457 MeV and 1.521 MeV.
There are, however, two data points for which the shell model result deviates from experiment:
for \(^{138}\textrm{Ba}\), the transition \(8_1^+ \to 6_1^+\) has an experimental value of \(0.32^{+0.79}_{-0.17}\) W.u. compared to the shell model value of 3.712 W.u., 
and the transition \(8_1^+ \to 6_2^+\) has an experimental value of \(0.18^{+0.47}_{-0.10}\) W.u. compared to the shell model value of 3.011 W.u.

For \(^{139}\textrm{La}\), the experimental \(B(E2;9/2_1^+ \to 7/2_1^+)\) value is \(7.4 \pm 1.1\) W.u., while the shell model predicts only $2.31 \times 10^{-5}$ W.u.
This data point lies outside the range shown in \figref{E2_log}.
Interestingly, the calculated \(B(E2;9/2_2^+ \to 7/2_1^+)\) is 7.382 W.u., close to the observed \(9/2_1^+ \to 7/2_1^+\) value. 
Similarly, the experimental \(9/2_1^+ \to 5/2_1^+\) transition is \(1.76 \pm 0.24\) W.u., while the shell model predicts 7.679 W.u. for this transition and {0.055} W.u. for \(9/2_2^+ \to 5/2_1^+\).
These results suggest a possible reversal in the calculated ordering of the \(9/2^+\) states.

\begin{figure}[h]
		\centering
		\includegraphics[width=0.9\linewidth]{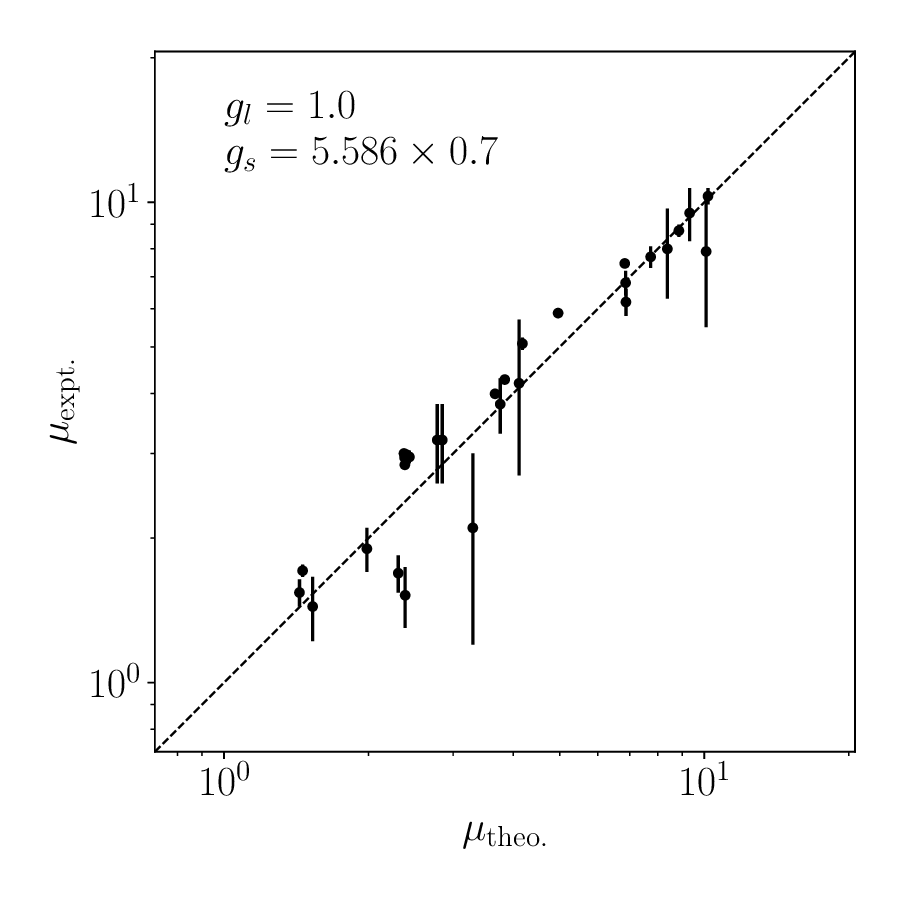}
		\caption{ Same as \figref{E2_log}, but for the magnetic moment $\mu$ (in $\mu_{\rm N}$). The experimental values are taken from Refs.~\cite{A=135,A=137,A=139,A=141,A=143,A=145,A=147,A=134,A=136,A=138,A=140,A=142,A=144,A=146}.
        }
		\label{gfactor}
\end{figure}

	\begin{figure*}
		\centering
		\includegraphics[width=0.9\linewidth]{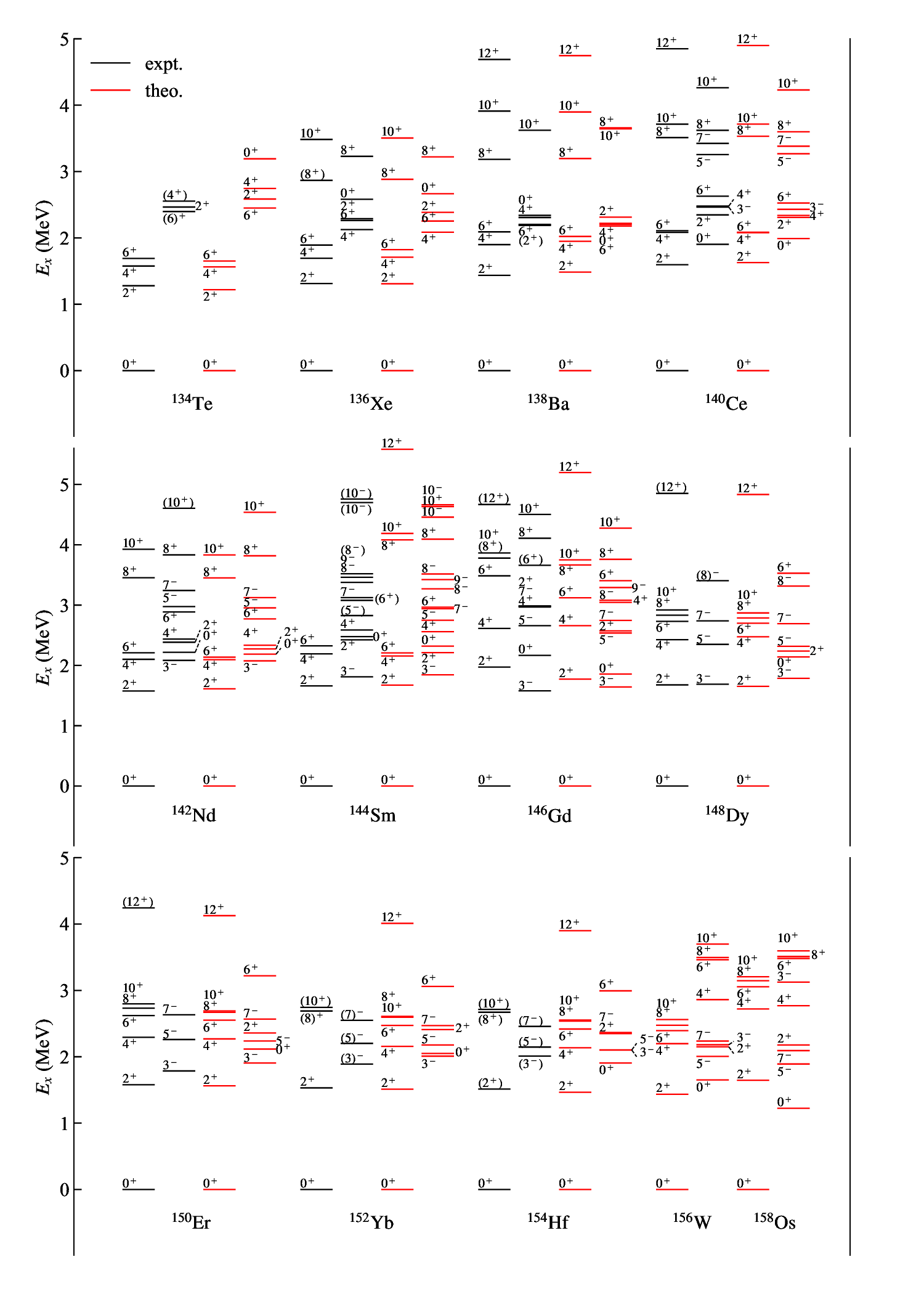}
		\caption{Low-lying spectra in the even-even isotones. The experimental values are taken from Refs.~\cite{A=134,A=136,A=138,A=140,A=142,A=144,A=146,A=148,A=150,A=152,A=154}.
        Only part of low-lying states are shown for simplicity.
        }
		\label{N82_even}
	\end{figure*}

The magnetic dipole moment, \(\mu\), is defined by
	\begin{equation}
			\mu \equiv \bra {J ; M=J} \sum_i g_l(i) \hat{l}_{z, i}+\sum_i g_s(i) \hat{s}_{z, i}\ket{J ; M=J} \mu_{\rm N} ,
	\end{equation}
where we take the values of the orbital and spin proton  \(g\)-factors \(g_{l} = 1.0\) and \(g_{s} = 5.586 \times 0.7\), with $0.7$ being the standard quenching factor. 
Fig.~\ref{gfactor} compares the calculated magnetic moments with 29 experimental data.
For reference, the corresponding experimental \(g\)-factors are listed in Table \ref{tablegfactor} in Appendix~\ref{appa}.
The shell model results reproduce the magnetic moments well, with most discrepancies smaller than $0.5$ $\mu_{\rm N}$.

\subsection{Low-lying structures of even-even nuclei}
	
\begin{figure}[H]
		\centering
		\includegraphics[width=1\linewidth]{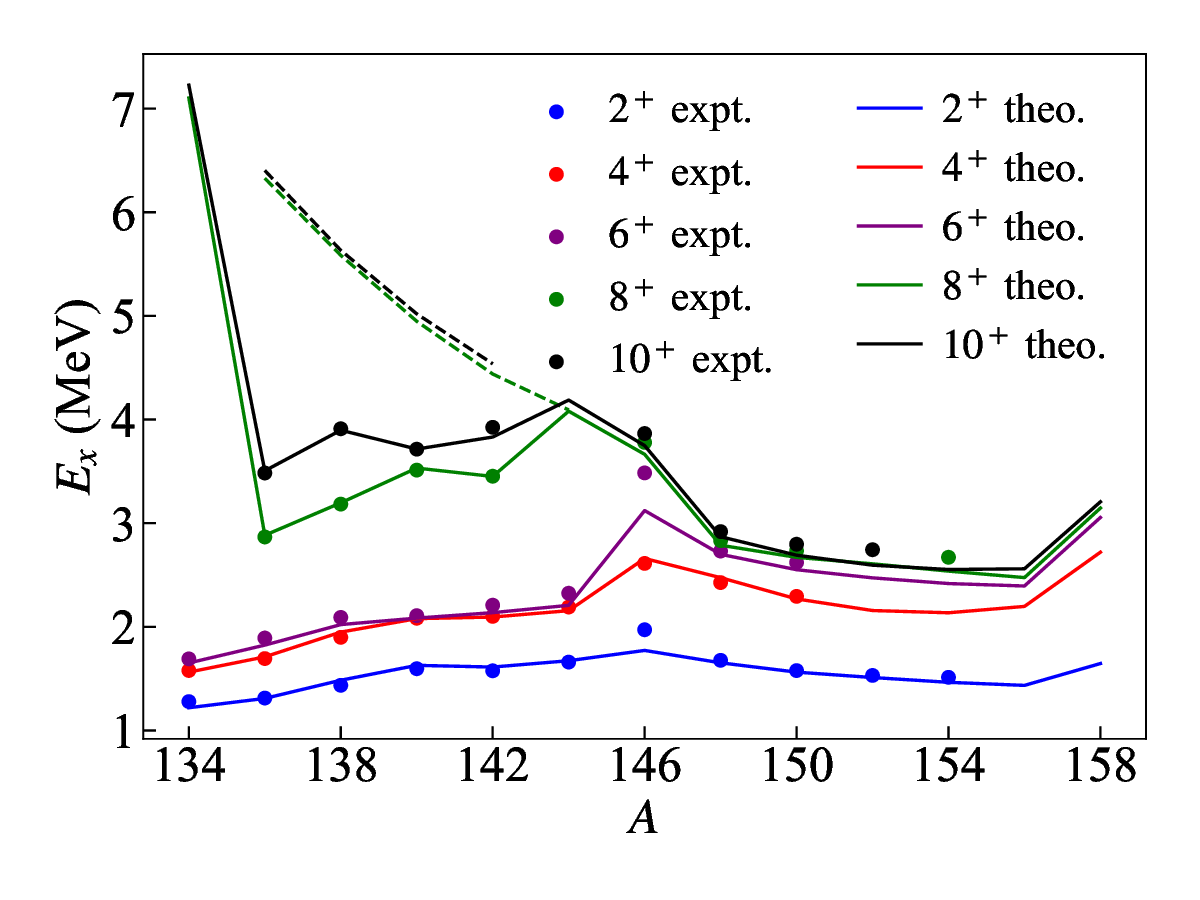}
		\caption{Systematics of the yrast 2\(^+\), 4\(^+\), 6\(^+\), 8\(^+\) and 10\(^+\) states in the even-even isotones (dots for experiment and solid curves for calculation). 
        The figure also shows (as dashed curves) the calculated non-yrast 8\(^+\) and 10\(^+\) states in \(^{136}\textrm{Xe}\)-\(^{142}\textrm{Nd}\), as well as the second 8\(^+\) state in \(^{144}\textrm{Sm}\), which are expected to be dominated by seniority-2 configurations (see discussion on \(g\)-factors in the main text). 
        }
		\label{246810}
\end{figure}

In this subsection, we discuss the low-lying states of the $N=82$ even-even nuclei.
Fig.~\ref{N82_even} presents selected spectra up to 5 MeV; not all calculated or experimental states are shown.
We begin with the systematics of the first $2^+$ state.
As shown in Fig.~\ref{246810}, the calculated excitation energies of the \(2_1^+\) states agree well with the experimental data. 
Experimentally, the \(2^+_1\) excitation energy reaches a pronounced peak at \(^{146}\textrm{Gd}\), with a value of 1.972 MeV, along the $N=82$ isotonic chain, indicating the presence of the $Z=64$ subshell closure~\cite{PhysRevLett.41.289,PhysRevLett.47.1433,PhysRevLett.85.720,g5pj-bngt}.
A weaker local maximum is also seen at \(^{140}\textrm{Ce}\), where the excitation energy is 1.596 MeV, only 20 keV higher than in the neighboring \(^{142}\textrm{Nd}\).
Our shell model calculations reproduce both features, with calculated excitation energies of 1.628 MeV for \(^{140}\textrm{Ce}\) and 1.772 MeV for \(^{146}\textrm{Gd}\).

\begin{figure}
		\centering
		\includegraphics[width=1\linewidth]{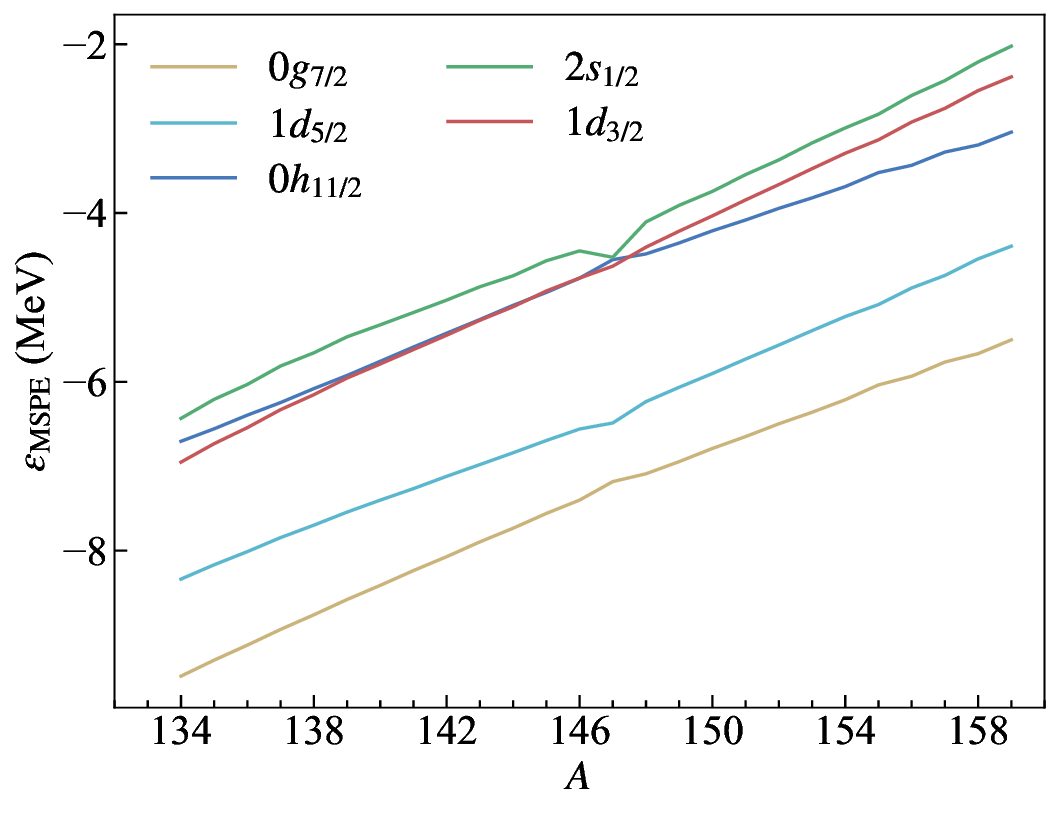}
		\caption{
        The mean-field single-particle energies (MSPEs) for the ground state of the $N=82$ isotones. 
        }
		\label{ESPE}
\end{figure}

To quantify the $Z=64$ subshell gap, we calculate the mean-field single-particle energies (MSPEs) derived from the monopole interaction, defined as
\begin{equation}
\varepsilon_{\mathrm{MSPE}}(a) = \varepsilon_a + \frac{1}{2} \sum_b  V_m(a b) \left[ \braket{\hat{n}(b)} - \delta_{ab} \right],
\end{equation}
where $\varepsilon_a$ is the SPE of orbital $a$, $\braket{\hat{n}(b)}$ is the occupation number of orbital $b$ obtained from the shell model calculation, and $V_m(a b)$ is the monopole matrix element of the two-body interaction \cite{ZUKER199465,PhysRevC.54.1641,RevModPhys.77.427}. In this work, we consider only the proton-proton monopole matrix elements, i.e.,
\begin{equation}
V_m(a b) \equiv \frac{ \sum_{J}  (2J+1) \left[ 1 - (-)^{J+T} \delta_{ab}\right] V_{J T}(a b a b)}{(2j_a+1)\left[(2j_b+1) + (-)^T \delta_{ab} \right]},
\end{equation}
with $T=1$.
The MSPE accounts for the spherical mean-field potential felt by each valence nucleon, arising from interactions with all others, in addition to the SPEs originating from the one-body potential of the inert core.
It is worth noting that the MSPE defined here is similar, though not identical, to the effective single-particle energy defined in Refs.~\cite{Otsuka_2013,RevModPhys.92.015002} (see Appendix~\ref{appb} for discussion).
The calculated MSPEs for the $0g_{7/2}, 1d_{5/2}, 1d_{3/2}, 2s_{1/2}$, and $0h_{11/2}$ orbitals, evaluated using the occupation numbers of the ground states in the $N=82$ isotones, are shown in Fig.~\ref{ESPE}. 
We find a subshell gap of {1.788} MeV at $Z=64$ between the $1d_{5/2}$ and $0h_{11/2}$ orbitals, which leads to the large \(2^+_1\) excitation energy in \(^{146}\textrm{Gd}\). Similarly, there is a gap of {1.062} MeV at $Z=58$ between the $0g_{7/2}$ and $1d_{5/2}$ orbitals, which explains the large \(2^+_1\) excitation energy in \(^{140}\textrm{Ce}\).

\begin{figure}
		\centering
		\includegraphics[width=1\linewidth]{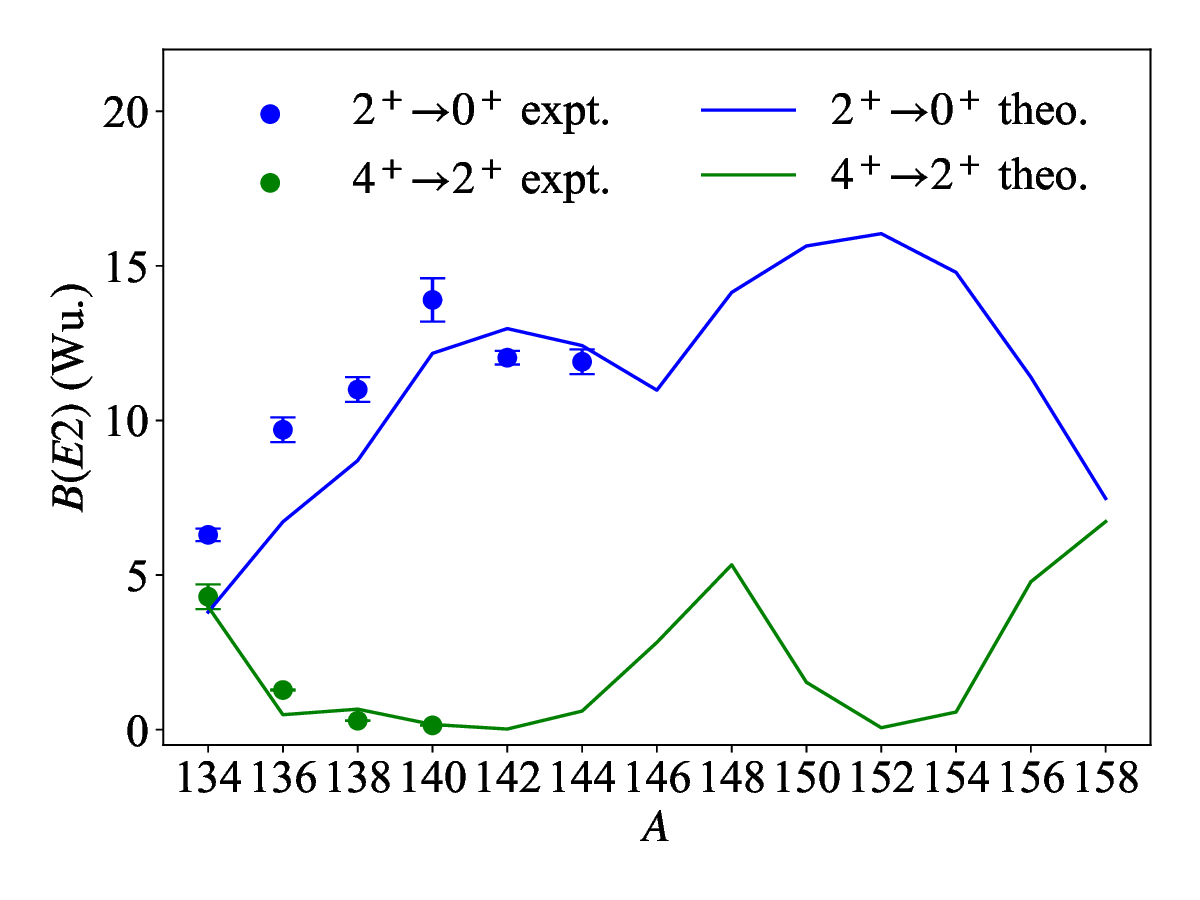}
		\caption{
        Systematics of the \(B(E2; 2_1^+ \to 0_1^+)\) and \(B(E2; 4_1^+ \to 2_1^+)\) values.
        }
		\label{BE024}
\end{figure}

The behavior of the \(B(E2; 2_1^+ \to 0_1^+)\) and \(B(E2; 4_1^+ \to 2_1^+)\) transition strengths is interesting.
In the single-$j$ seniority scheme~\cite{PhysRev.63.367,FLOWERS19521101,racah1952nuclear}, the \(0_1^+\) state has seniority \(\nu  = 0\), while the \(2_1^+\), \(4_1^+\), and \(6_1^+\) states typically have seniority \(\nu  = 2\).
In this scenario, \(B(E2)\) strengths with \(\Delta \nu  = 2\) (e.g., \(2_1^+ \to 0_1^+\)) follow a smooth parabolic trend with the valence nucleon number, reaching a maximum at midshell. 
Extensive shell model and generalized seniority calculations for semi-magic nuclei produced this trend.
A well-known exception occurs in the Sn isotopes, where the experimental \(B(E2; 2_1^+ \to 0_1^+)\) values exhibit a shallow minimum at \(^{116}\textrm{Sn}\), which has motivated numerous experimental and theoretical investigations~\cite{PhysRevLett.101.012502,JUNGCLAUS2011110,MORALES2011606,PhysRevC.72.061305,ANSARI200537,PhysRevC.74.054313,TERASAKI2004583,PhysRevC.84.044314,PhysRevC.86.054304, PhysRevLett.121.062501}.
Our shell model calculations for the $N=82$ isotones show a similar deviation: the calculated \(B(E2; 2_1^+ \to 0_1^+)\) value exhibits a shallow minimum at \(^{146}\textrm{Gd}\) (see \figref{BE024}), which may be attributed to the $Z=64$ subshell closure.
This indicates that such a phenomenon can emerge within a one-major-shell description using an optimized interaction, without invoking large-scale excitations across $N,Z=50$.
For the \(B(E2; 4_1^+ \to 2_1^+)\) values, the single-$j$ seniority scheme predicts a parabolic trend with a minimum at midshell, whereas our shell model calculations yield instead a local maximum at \(^{148}\textrm{Dy}\).

As shown in Figs.~\ref{N82_even} and \ref{246810}, the even-even nuclei with $Z<64$ (\(^{134}\textrm{Te}\)-\(^{144}\textrm{Sm}\)) exhibit a small energy difference between the \(4_1^+\) and \(6_1^+\) states.
This leads to isomerism of the \(6_1^+\) states, with measured half-lives ranging from 0.163 $\mu$s to 16.5 $\mu$s, which are consistently interpreted as seniority-2 isomers.

\begin{figure}[h]
		\centering
		\includegraphics[width=1\linewidth]{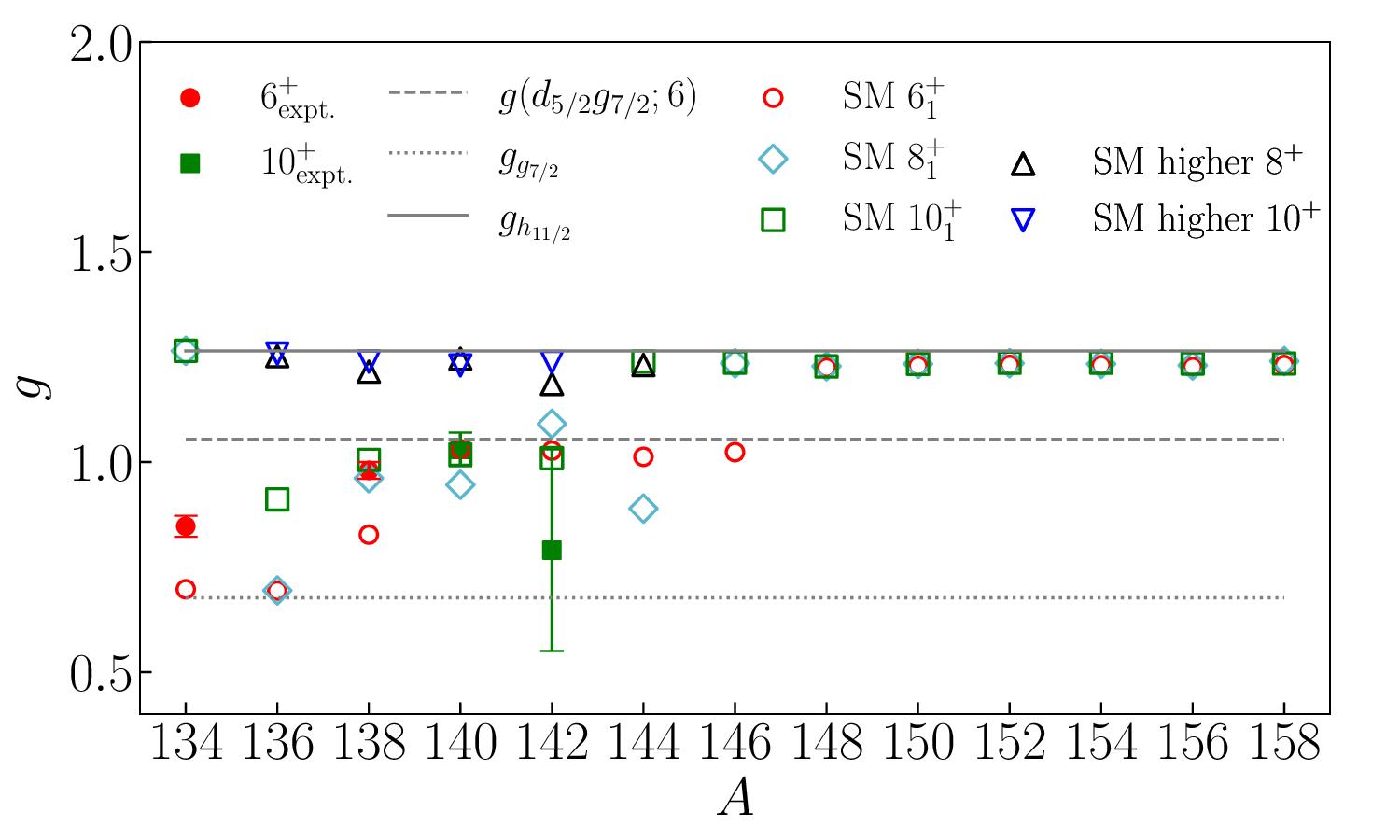}
			\caption{
The \(g\)-factor of the yrast 6\(^+\), 8\(^+\), and 10\(^+\) states in the even-even isotones. 
The solid points are the experimental data, the open points represent the shell model result, and the lines denote the two-nucleon estimate $g(j_1j_2;J)$ defined in \eqref{mu-s2}.
The figure also shows the calculated \(g\)-factors for non-yrast 8\(^+\) and 10\(^+\) states in \(^{136}\textrm{Xe}\)-\(^{142}\textrm{Nd}\), and the second 8\(^+\) state in \(^{144}\textrm{Sm}\), which are expected to be dominated by seniority-2 configurations (see discussion in the text). 
}
		\label{g-factor}
	\end{figure}

We next analyze the magnetic \(g\)-factors of the \(6_1^+\) states to further probe their seniority structure.
The \(g\)-factor is defined by
\begin{equation}
   	 g = \frac{\mu}{J\mu_{\rm N}},
\end{equation}
where \(J\) is the total spin of the state.
For a single nucleon, the Schmidt value of the \(g\)-factor is
\begin{equation}\label{mu-s}
		\begin{aligned}
			g_j
			=\begin{cases}
				 g_l +\frac{ g_s-g_l  }{2j},   \quad ~\text{if } j=l+\frac{1}{2}, \\
				g_l  -\frac{ g_s-g_l  }{2(j+1)} ,   \quad \text{if } j=l-\frac{1}{2}.
			\end{cases}
		\end{aligned}
	\end{equation}
For two nucleons in the $j_1$ and $j_2$ orbitals coupled to $J$, the \(g\)-factor can be further derived from the Schmidt values as follows: 
\begin{equation}\label{mu-s2} 
\begin{aligned}
g(j_1j_2;J) =~& \frac{j_1(j_1 +1)  - j_2(j_2 +1)  }{2J(J +1)}(g_{j_1} - g_{j_2})  \\
& +\frac{1}{2}(g_{j_1} + g_{j_2}) .
\end{aligned}
\end{equation}
It is worth noting that \eqref{mu-s2} offers an approximate estimate of \(g\)-factors in an even-even nucleus $A$, valid under the assumption of weak coupling between the two nucleons and the $A-2$ core (e.g., seniority-2 states)~\cite{PhysRev.122.1530,Gerda_Neyens_2003}. This formula also neglects antisymmetrization between the core and the two nucleons.
It can be seen that the value of $g(j_1j_2;J)$ is reduced to the Schmidt value $g_j$ if $j_1 = j_2$. 
For the \(6_1^+\) states, we calculate $g(j_1j_2;J)$ for the two-nucleon configurations $(0g_{7/2})^2$, $0g_{7/2}1d_{5/2}$, and $(0h_{11/2})^2$.
The results are shown in Fig.~\ref{g-factor}.
For the \(6_1^+\) states in \(^{134}\textrm{Te}\) and \(^{136}\textrm{Xe}\), the shell model yields $g=0.697$ and $0.694$, respectively, which are very close to the Schmidt value $g_{g_{7/2}} = 0.677$.
This indicates that these two \(6_1^+\) states are dominated by the seniority-2 $(0g_{7/2})^2$ configuration.
The shell model also gives occupation numbers for the $0g_{7/2}$ orbital of 1.945 and 3.573,
respectively, further supporting this assignment.
Similarly, for \(^{140}\textrm{Ce}\), \(^{142}\textrm{Nd}\), \(^{144}\textrm{Sm}\), and \(^{146}\textrm{Gd}\), the shell model \(g\)-factors indicate that the \(6_1^+\) state is dominated by the seniority-2 $0g_{7/2}1d_{5/2}$ configuration (see Fig.~\ref{g-factor}).
The highest excitation energy of the \(6_1^+\) state occurs at \(^{146}\textrm{Gd}\), and the shell model result is in good agreement with this phenomenon.
For \(\mathrm{^{148}Dy}\), \(\mathrm{^{150}Er}\), \(\mathrm{^{152}Yb}\), and \(\mathrm{^{154}Hf}\) ($Z>64$), the \(6_1^+\) states correspond to the $(0h_{11/2})^2$ configuration.


Our shell model calculation also reproduces the systematic behavior of the yrast \(8^+\) and \(10^+\) states (see Fig.~\ref{246810}). For \(^{134}\textrm{Te}\), the \(8_1^+\) and \(10_1^+\) states have not yet been identified experimentally, but the shell model predicts their excitation energies to be 7.102 MeV and 7.226 MeV, respectively. 
These two states correspond to the \((0h_{11/2})^2\) configuration.
In \(^{136}\textrm{Xe}\), \(^{138}\textrm{Ba}\), \(^{140}\textrm{Ce}\), and \(^{142}\textrm{Nd}\), the energy gap between the yrast \(8^+\) and \(10^+\) states is relatively large and does not follow the typical seniority pattern.
Consistently, the calculated \(g\)-factors for these states deviate from the Schmidt value $g_{h_{11/2}} = 1.265$. 
Due to the very high excitation energies predicted for the \(8_1^+\) and \(10_1^+\) states in \(^{134}\textrm{Te}\), we expect that the corresponding seniority-2 states in \(^{136}\textrm{Xe}\)-\(^{142}\textrm{Nd}\) also have high excitation energies.
Indeed, our shell model predicts the seniority-2 \(8^+\) and \(10^+\) states at excitation energies of 6.327/6.402 MeV in \(^{136}\textrm{Xe}\), 5.584/5.632 MeV in \(^{138}\textrm{Ba}\), 4.948/5.019 MeV in \(^{140}\textrm{Ce}\), and 4.439/4.538 MeV in \(^{142}\textrm{Nd}\), respectively. As shown in Fig.~\ref{g-factor}, the \(g\)-factors for these states are close to $1.265$ for the \((0h_{11/2})^2\) configuration, confirming their seniority-2 nature.
For \(^{144}\textrm{Sm}\), the \(8^+\) and \(10^+\) states have not yet been clearly identified experimentally. The shell model predicts two low-lying \(8^+\) states with very similar energies: 4.080 MeV and 4.093 MeV. 
According to Fig.~\ref{g-factor}, the \(8_1^+\) state does not exhibit seniority-2 features, whereas the \(8_2^+\) state is dominated by the seniority-2 \((0h_{11/2})^2\) configuration.
The predicted \(10_1^+\) state in  \(^{144}\textrm{Sm}\) also has a dominant seniority-2 character.
With modern high-purity germanium detectors achieving gamma-ray energy resolutions of 1-2 keV, these two nearly degenerate $8^+$ states may be resolved in future experiments.
For heavier isotones such as \(^{146}\textrm{Gd}\), \(^{148}\textrm{Dy}\), \(^{150}\textrm{Er}\), \(^{152}\textrm{Yb}\), \(^{154}\textrm{Hf}\), \(^{156}\textrm{W}\), and \(^{158}\textrm{Os}\), the \(8^+\) and \(10^+\) states exhibit a consistent \((0h_{11/2})^2\) character.

\figref{246810} further reveals a clear systematic trend for the excitation energies of these seniority-2 \(8^+\) and \(10^+\) states across the $N=82$ isotones. 
For $Z \leq 64$, the excitation energies decrease with increasing proton number, as the proton Fermi surface approaches the $0h_{11/2}$ orbital. 
This trend gradually saturates as $Z$ approaches 64.
Interestingly, at \(Z=74\), the excitation energies of the \(2^+\), \(4^+\), \(6^+\), \(8^+\), and \(10^+\) states begin to rise again. 
This behavior is attributed to the approximately full occupancy of the \(0g_{7/2}\), \(1d_{5/2}\), and \(0h_{11/2}\) orbitals in the ground state of \(^{158}\textrm{Os}\).
The enhanced excitation energies in the excited states arise from the energy gap between the \(0h_{11/2}\) and \(1d_{3/2}\) orbitals at \(Z=74\) (see \figref{ESPE}).

For the non-yrast states, we focus on the excitation energies of the \(0_2^+\) states. 
The \(0_2^+\) states have been experimentally observed in \(^{136}\textrm{Xe}\), \(^{138}\textrm{Ba}\), \(^{140}\textrm{Ce}\), \(^{142}\textrm{Nd}\), and \(^{144}\textrm{Sm}\), and our shell model calculations are close to the experimental values. 
However, for \(^{146}\textrm{Gd}\), the experimental excitation energy of the \(0_2^+\) state is 2.165 MeV, while the shell model predicts 1.875 MeV, resulting in a difference of 0.308 MeV. 
The excitation energies of the \(0_2^+\) states in \(^{136}\textrm{Xe}\)-\(^{146}\textrm{Gd}\) vary slowly during our iterative PCA fitting procedure, indicating that they are not very sensitive to the proton-proton TBMEs in the $g_7dsh_{11}$ shell space.

\begin{figure}[h]
		\centering
		\includegraphics[width=1\linewidth]{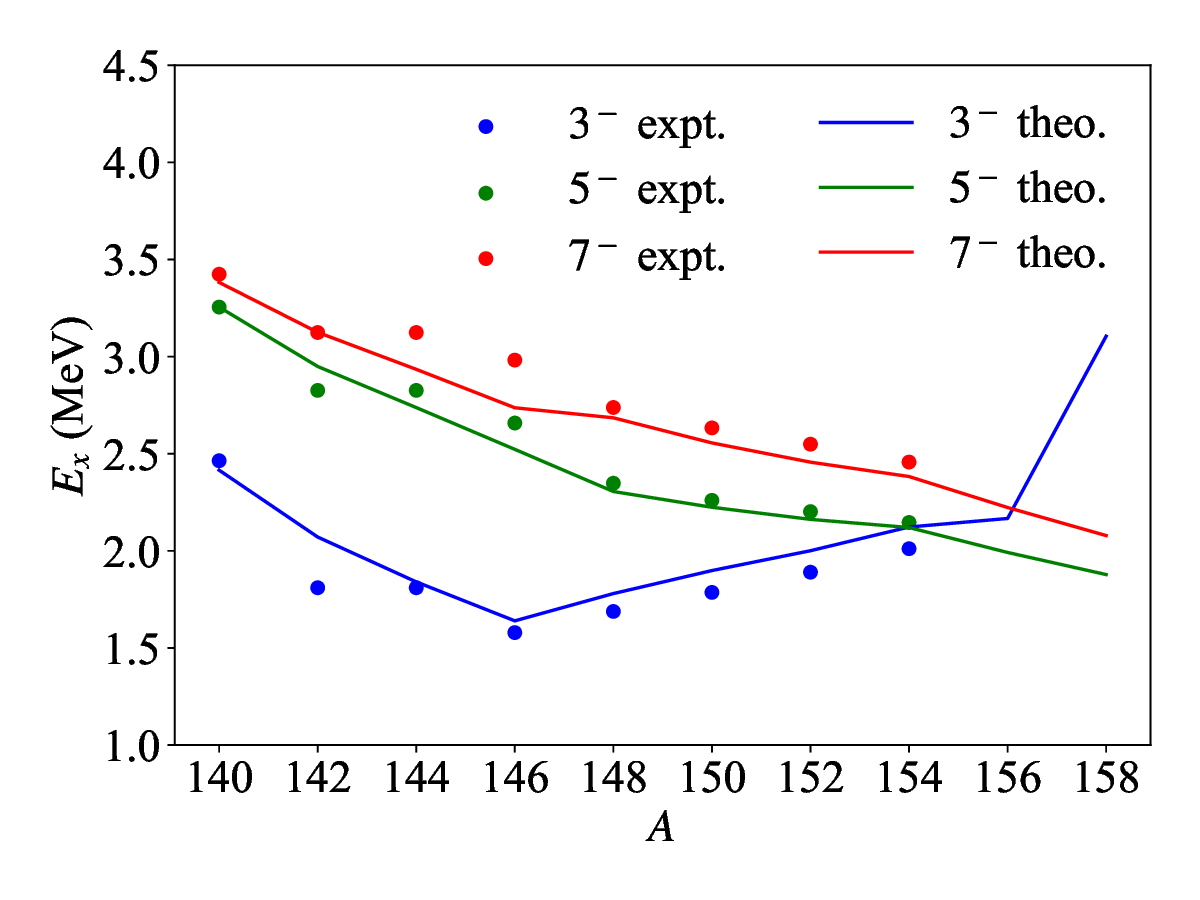}
		\caption{Systematics of the excitation energies for the first 3\(^-\), 5\(^-\), and 7\(^-\) states. }
		\label{357}
\end{figure}

Now let us turn to the low-lying odd-parity states. 
Previous shell model interactions~\cite{BROWN2014115,kuo1971us,PhysRevC.43.602,PhysRevC.45.1720,PhysRevC.80.044320} lead to noticeable discrepancies when compared to experimental results for the yrast \(3^-\), \(5^-\), and \(7^-\) states. In particular, they generally overestimate the \(3^-\) excitation energies.
The systematics of the excitation energies for the \(3_1^-\), \(5_1^-\), and \(7_1^-\) states are shown in Fig.~\ref{357}. 
Overall, our result shows good agreement with the experimental values. 
In particular, there is a pronounced minimum in the \(3_1^-\) state of \(^{146}\textrm{Gd}\), which is well reproduced by the shell model calculation. 
This feature can be understood as follows: for \(^{146}\textrm{Gd}\), the proton Fermi surface lies between the $1d_{5/2}$ and $0h_{11/2}$ orbitals. The $1d_{5/2}$-$0h_{11/2}$ coupling is energetically favorable, producing the lowest \(3_1^-\) state among the $N=82$ isotones.
The calculated \(g\)-factor of the \(3^-\) state in \(^{146}\textrm{Gd}\) is 1.099, consistent with the two-nucleon estimate \(g(h_{11/2}d_{5/2};3) = 1.066\). 
For the other nuclei, the shell model \(g\)-factors of the \(3^-\) states range from 1.088 to 1.104, also close to \(g(h_{11/2} d_{5/2}; 3) \), highlighting the dominant role of this coupling near the $Z=64$ subshell.



We also investigate the \(B(E3; 3^- \rightarrow 0^+)\) transition probabilities. 
Experimentally, the \(B(E3)\) value for the \(3^- \to 0^+\) transition in \(^{140}\textrm{Ce}\) is reported as \(26^{+12}_{-18}\) W.u., while our shell model calculation yields 8.681 W.u. 
For \(^{146}\textrm{Gd}\), the experimental value is 37(5) W.u., and the shell model result is 18.29 W.u.
These results indicate that the shell model underestimates the \(B(E3)\) strengths by approximately a factor of two with an effective proton charge of $e_{\pi} = 1.5e$.
This discrepancy is likely due to missing contributions from neutron excitations across the $N=82$ major shell gap~\cite{HOLT1997107}.
If an enlarged effective charge of 2.0 is adopted, the shell model \(B(E3)\) values in \(^{140}\textrm{Ce}\) and \(^{146}\textrm{Gd}\) become 15.433 and 32.516 W.u., respectively, both of which are consistent with the experimental results within the uncertainties.

\begin{figure*}
		\centering
		\includegraphics[width=0.9\linewidth]{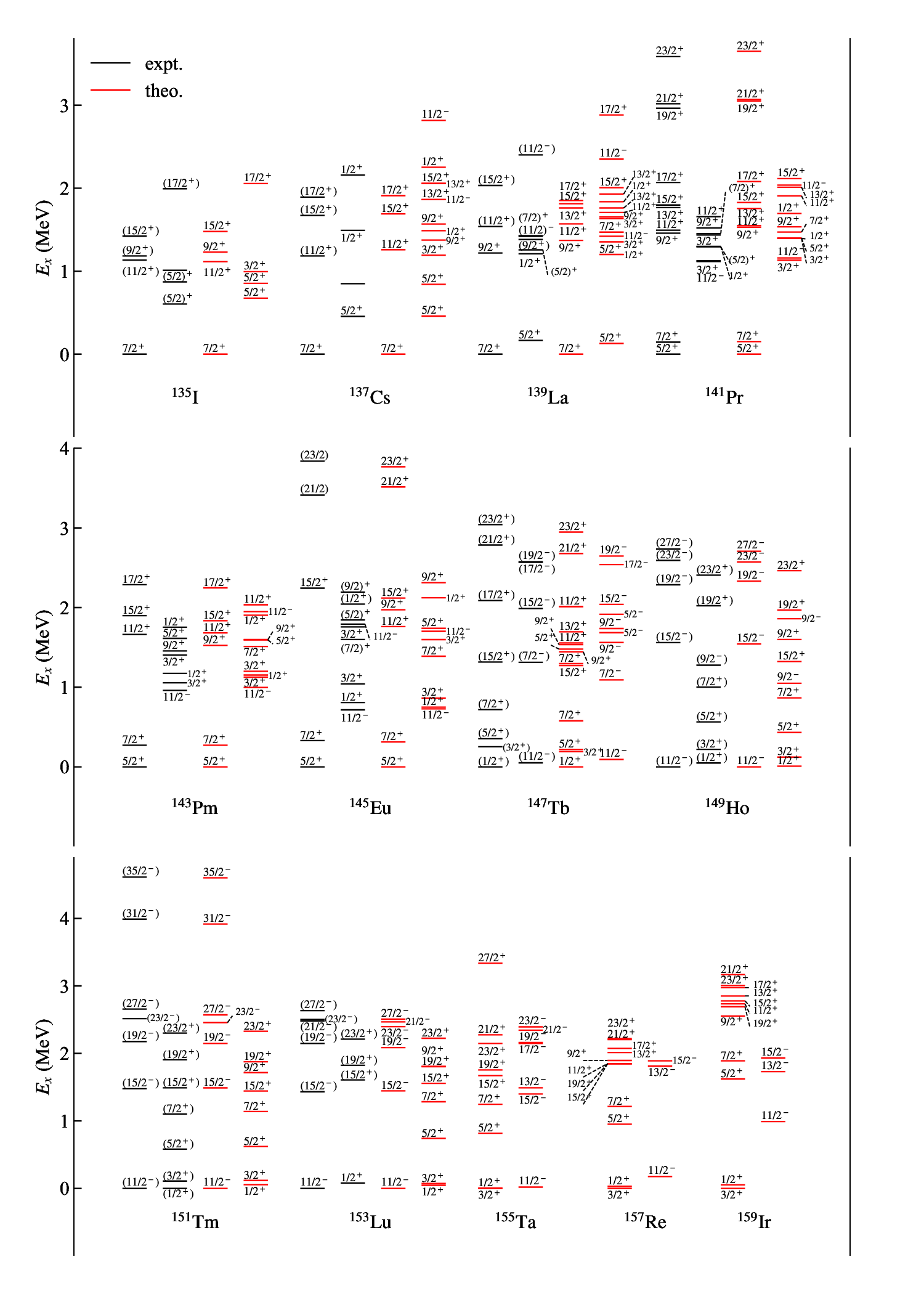}
		\caption{Low-lying spectra in the odd-$A$ isotones. The experimental values are taken from Refs.~\cite{A=135,A=137,A=139,A=141,A=143,A=145,A=147,A=149,A=151,A=153}.
        } 
		\label{N82_odd}
\end{figure*}

\subsection{Low-lying structures of odd-\(A\) nuclei}

In this section, we study the low-lying structures of the $N=82$ odd-\(A\) nuclei. The low-lying spectra are shown in Fig.~\ref{N82_odd}. Overall, the shell model results for excitation energies are in good agreement with the experimental data. 

Our calculations correctly reproduce the ground-state spins and parities across the $N=82$ isotonic chain. 
In comparison, previous shell model interactions generally fail, except for CW5082 \cite{PhysRevC.45.1720} and the interaction in Ref.~\cite{PhysRevC.80.044320}, which correctly describe the spins and parities for the isotones with $Z \leq 65$.
For $Z < 64$, the ground-state systematics can be understood from the MSPEs.
As shown in \figref{ESPE}, the gap between the \(0g_{7/2}\) and \(0d_{5/2}\) orbitals and the $Z=64$ subshell closure separating the \(0d_{5/2}\) and \(0h_{11/2}\) orbitals underlie the systematics.
Accordingly, the ground states of \(^{135}\textrm{I}\), \(^{137}\textrm{Cs}\), and \(^{139}\textrm{La}\) are \(7/2^+\), reflecting the dominance of the \(0g_{7/2}\) orbital. 
The ground states of \(^{141}\textrm{Pr}\), \(^{143}\textrm{Pm}\), and \(^{145}\textrm{Eu}\) are \(5/2^+\), attributed to the filling of the \(0g_{7/2}\) orbital and the onset of occupancy in the \(0d_{5/2}\) orbital.

For $Z > 64$, theoretical predictions of the ground-state spins and parities become more challenging.
For example, the \(1/2^+\) ground state of \(^{147}\textrm{Tb}\) and the \(3/2^+\) ground states of \(^{155}\textrm{Ta}\) and \(^{157}\textrm{Re}\) cannot be directly explained by the SPEs or MSPEs shown in Fig.~\ref{ESPE}, because the MSPEs display near degeneracy among the \(1d_{3/2}\), \(2s_{1/2}\), and \(0h_{11/2}\) orbitals, and residual two-body interactions beyond the spherical mean field play a critical role here.
Our shell model calculations reproduce the data.
For comparison, the \(11/2^-\) ground states of \(^{149}\textrm{Ho}\), \(^{151}\textrm{Tm}\), and \(^{153}\textrm{Lu}\) and the \(3/2^+\) ground state of \(^{159}\textrm{Ir}\) are consistent with the mean-field scenario.



Interestingly, the shell model predicts several very low-lying excited states in the odd-$A$ isotones, suggesting the possible isomerism.
For instance, a \(1/2^+\) state is predicted in \(^{155}\textrm{Ta}\) with extremely low excitation energy of $E_x = 4$ keV!
Similarly, low-lying \(1/2^+\) states are predicted in \(^{151}\textrm{Tm}\), \(^{157}\textrm{Re}\), and \(^{159}\textrm{Ir}\) at {55 keV, 30 keV, and 52 keV}, respectively.
In addition, the shell model predicts a \(3/2^+\) state in \(^{153}\textrm{Lu}\) at {72 keV}, and an \(11/2^-\) state in \(^{155}\textrm{Ta}\) at {19 keV}.

Now let us discuss the configurations of very low-lying states. 
The \(11/2_1^-\) states in all 13 odd-$A$ isotones correspond to the seniority-1 \(0h_{11/2}\) configuration, as indicated by their shell model \(g\)-factors, which are very close to the Schmidt value $g_{h_{11/2}} = 1.265$.
The ground states of \(^{135}\textrm{I}\), \(^{137}\textrm{Cs}\), and \(^{139}\textrm{La}\) are \(7/2^+\), with calculated \(g\)-factors of 0.679, 0.680, and 0.685, respectively. 
These values are close to $g_{g_{7/2}} = 0.677$ and indicate a nearly pure \(0g_{7/2}\) single-particle character.
For the \(7/2_1^+\) states in the heavier isotones \(^{141}\textrm{Pr}\), \(^{143}\textrm{Pm}\), \(^{145}\textrm{Eu}\), \(^{147}\textrm{Tb}\), \(^{149}\textrm{Ho}\), \(^{151}\textrm{Tm}\), \(^{153}\textrm{Lu}\), \(^{155}\textrm{Ta}\), \(^{157}\textrm{Re}\), and \(^{159}\textrm{Ir}\), the \(g\)-factors gradually increase, with the values of 0.695, 0.707, 0.719, 0.717, 0.719, 0.739, 0.839, 0.919, 0.932, and 0.921, respectively. 
This trend reflects a gradual increase in configuration mixing with increasing excitation energy.

The structure of the low-lying \(5/2^+\) states is more complex. 
In \(^{135}\textrm{I}\), two low-lying states with tentative spin-parity assignments of \(5/2^+\) have been observed at 0.604 MeV and 0.871 MeV. 
Their spins are not firmly established, and thus they are not included in our PCA fitting. 
Our shell model predicts two \(5/2^+\) states at 0.674 MeV and 0.852 MeV, respectively. 
Previous work~\cite{SUHONEN199841} suggested that the \(5/2_1^+\) state is a single-particle state, while the \(5/2_2^+\) state corresponds to a weak-coupling state $\ket{ (2^+ \times 0g_{7/2})^{(5/2^+)} }$.
Our shell model result, however, shows that both states are admixtures. 
The calculated \(g\)-factor of the lower state is 0.969, significantly smaller than $g_{d_{5/2}} = 1.582$, while the \(g\)-factor of the upper state is 1.277. 
To interpret this, we estimate the \(g\)-factor of a weak-coupling configuration $\ket{ (2_1^+ \times j)^{(J)} }$ approximately by
\begin{equation} \label{seniority3}
\begin{aligned}
& g\left[ (2_1^+ \times j)^{(J)} \right] \\
 & \quad \approx  \frac{1}{2 J(J+1)}\left\{ ~ g(2_1^{+}) \left[ 6+J(J+1)-j(j+1) \right] \right. \\
& \qquad\qquad\qquad\quad~ \left. + g_j \left[ j(j+1)+J(J+1)-6 \right] ~ \right\} \\
&  \quad   =0.676, \quad\text{for } j = 7/2 \text{ and } J = 5/2,
\end{aligned}
\end{equation}
where we used $g(2_1^+)=0.691$ from the shell model for \(^{134}\textrm{Te}\).
Similar to \eqref{mu-s2}, \eqref{seniority3} provides an approximate \(g\)-factor for seniority-3 states in odd-$A$ nuclei, assuming weak coupling and neglecting antisymmetrization.
Comparing \eqref{seniority3} with the shell model \(g\)-factors, we estimate that the \(5/2_1^+\) state in \(^{135}\textrm{I}\) is approximately a 70\% $\ket{ (2^+ \times 0g_{7/2})^{(5/2^+)} }$ configuration and 30\% single-particle $1d_{5/2}$ configuration, while the \(5/2_2^+\) state is primarily $1d_{5/2}$.
A similar analysis suggests that the \(3/2_1^+\) state of \(^{135}\textrm{I}\) is nearly a pure $\ket{ (2^+ \times 0g_{7/2})^{(3/2^+)} }$ weak-coupling state, as its shell model $g$-factor is 0.676.

In \(^{137}\textrm{Cs}\), the \(5/2_1^+\) state has a \(g\)-factor of 1.563 and the \(1d_{5/2}\) occupation increases from 0.61 in the \(7/2^+\) ground state to 1.37 in this excited state, confirming a seniority-1 character. 
The dominant configuration of the \(5/2_2^+\) state is $\ket{ (2^+ \times 0g_{7/2})^{(5/2^+)} }$, as the shell model $g$-factor is 0.683, close to the weak-coupling value in \eqref{seniority3}, and the \(0g_{7/2}\) occupation is 4.41.
The $g$-factors of the \(5/2_1^+\) states in the heavier isotones, \(^{139}\textrm{La}\), \(^{141}\textrm{Pr}\), \(^{143}\textrm{Pm}\), \(^{145}\textrm{Eu}\), \(^{147}\textrm{Tb}\), \(^{149}\textrm{Ho}\), \(^{151}\textrm{Tm}\), \(^{153}\textrm{Lu}\), \(^{155}\textrm{Ta}\), and \(^{157}\textrm{Re}\), are 1.557, 1.537, 1.504, 1.467, 1.481, 1.490, 1.494, 1.493, 1.483, and 1.459, respectively. 
Although these values show a slight downward trend with increasing $Z$, they remain close to the Schmidt value, confirming the predominantly seniority-1 character.
In \(^{159}\textrm{Ir}\), however, the $g$-factor of \(5/2_1^+\) drops significantly to 0.967 and the excitation energy is much higher than \(5/2_1^+\) in the lighter isotones, indicating stronger configuration mixing or structural changes in this heavy nucleus.

\section{summary and outlook}

In this work, we have performed comprehensive shell model calculations for the \(N=82\) isotonic chain, based on a newly optimized proton-proton effective interaction.
We use the PCA method to optimize the effective interaction by fitting to 204 experimental data points.  We consider in the optimization up to 30 degrees of freedom for linear combinations among the 165 TBMEs and SPEs of the effective Hamiltonian, which allows us to achieve a root-mean-square deviation of 102 keV. 
The resulting interaction successfully reproduces binding energies, as well as one- and two-proton separation energies, electromagnetic properties, and low-lying spectra, and provides reliable predictions for nuclei near or beyond the proton drip line, including \(^{155}\mathrm{Ta}\), \(^{156}\mathrm{W}\), \(^{157}\mathrm{Re}\), \(^{158}\mathrm{Os}\), and \(^{159}\mathrm{Ir}\).

We have studied the \(B(E2)\) values and \(g\)-factors which can provide a sensitive probe for low-lying structures across the even-even isotones, with particular focus on the seniority pattern.
The systematics of odd-parity states (\(3^-\), \(5^-\), and \(7^-\)) were analyzed. The shell model calculations correctly reproduces the experimental minimum of the \(3^-\) energy at \(^{146}\textrm{Gd}\). 
We also calculated \(B(E3)\) values and compared them with the available data.

For the odd-\(A\) nuclei, our shell model calculation reproduces all known ground-state spins and parities and stresses the importance of residual two-body interactions in describing the \(1/2^+\) ground state of \(^{147}\textrm{Tb}\) and the \(3/2^+\) ground states of \(^{155}\textrm{Ta}\) and \(^{157}\textrm{Re}\), beyond the scope of the mean-field approximation. 
We present systematic study for \(g\)-factors of low-lying \(5/2^+\), \(7/2^+\), and \(11/2^-\) states, and analyzed their underlying configurations. 
In particular, we suggest that the low-lying \(5/2^+\) states in \(^{135}\textrm{I}\) exhibit strong admixtures of seniority-one and weak-coupling configurations.

The optimized proton-proton effective interaction in this work is provided as Supplemental Material \cite{ralp5a}.
We hope that this interaction provides a solid foundation for future studies of atomic nuclei near \(N=82\).
Given the several successful applications of this interaction, both completed and ongoing, we mention the study of nuclear shape evolution and quantum phase transitions in rare-earth nuclei~\cite{g5pj-bngt,ggxp-98gd}, systematic investigations of even-even $N=80$ isotones~\cite{lwnd-3tgg,tobepub1}, and the first spectroscopic measurement of \(^{150}\textrm{Yb}\)~\cite{tobepub2}.

\begin{acknowledgments}

The authors thank Prof. H. Jiang and Dr. W. Q. Zhang for valuable discussions.
This work is supported by the National Natural Science Foundation of China under Grant Nos.~12322506, 12535009, and 12075169,
the Fundamental Research Funds for the Central Universities under Grant No.~22120250299,
and the High Performance Computing platform at Tongji University.
CQ acknowledges support from the Olle Engkvist Foundation.
YXY acknowledges J. X. Feng, Y. Chen, and Y. N. Zhu for their assistance in verifying the data.

\end{acknowledgments}

\appendix

\onecolumngrid

\section{Tables of \(B(E2)\) values and \(g\)-factors} \label{appa}

Based on the overall good agreement between experimental data and the shell model calculations, we present comprehensive tables of $B(E2)$ values and \(g\)-factors, respectively, for low-lying states in the $N=82$ nuclei.
\newline

\twocolumngrid

\begin{longtable}{c|c|c|c} \label{BE2} \\
\caption{ \(B(E2)\) values (in W.u.) in low-lying states of \(N=82\) isotones.  } \\
    	\hline\hline Nuclide & $I_i^\pi \rightarrow I_f^\pi$ & $B(E 2)_{\text {expt. }}$   & $B(E 2)_{\text {theo. }}$  \\
		\hline $^{134}\mathrm{Te}$ & $2_1^{+} \rightarrow 0_1^{+}$ & 6.3(20) & 3.802 \\
		          & $4_1^{+} \rightarrow 2_1^{+}$ & 4.3(4) & 3.978 \\
		          & $6_1^{+} \rightarrow 4_1^{+}$ & 2.05(4) & 1.968 \\
                    & $8_1^{+} \rightarrow 6_1^{+}$ & - &   0.003\\
                    & $10_1^{+} \rightarrow 8_1^{+}$ & - &   1.192\\
            \hline ${ }^{135} \mathrm{I}$ & $5 / 2_1^{+} \rightarrow 7 / 2_1^{+} $ & -& 7.643 \\
            & $ 5 / 2_2^{+}  \rightarrow 7 / 2_1^{+}$ & -& 1.482\\
		\hline ${ }^{136} \mathrm{Xe}$ & $2_1^{+} \rightarrow 0_1^{+}$ & 9.7(4) & 6.722\\
		          & $4_1^{+} \rightarrow 2_1^{+}$ & 1.281(17) & 0.482 \\
		       & $6_1^{+} \rightarrow 4_1^{+}$ & 0.0132(8) & 0.013 \\
		         & $6_2^{+} \rightarrow 4_1^{+}$ & $>0.26$ & 0.091\\
                  & $8_1^{+} \rightarrow 6_1^{+}$ & - &  3.514\\
                   & $10_1^{+} \rightarrow 8_1^{+}$ & - &  0.167\\
		\hline ${ }^{137} \mathrm{Cs}$ & $ 5 / 2_1^{+} \rightarrow 7 / 2_1^{+} $ & $>6.8$ & 0.157 \\
		\hline ${ }^{138} \mathrm{Ba}$ & $2_1^{+} \rightarrow 0_1^{+}$ & 11.0(4) & 8.704 \\
		           & $4_1^{+} \rightarrow 2_1^{+}$ & $0.2878(15)$ & 0.662 \\
		           & $4_2^{+} \rightarrow 2_1^{+}$ & $2.0(+17-7)$ & 1.149\\
		           & $6_1^{+} \rightarrow 4_1^{+}$ & $0.053(+8-6)$ &0.177\\
		           & $8_1^{+} \rightarrow 6_1^{+}$ & $0.32(+79-17)$ & 3.712\\
		           & $8_1^{+} \rightarrow 6_2^{+}$ & $0.18(+47-10)$ & 3.011\\
		           & $10_2^{+}(3910.5) \rightarrow 8_1^{+}$ & $>4.1$ &1.821\\
		           & $10_1^{+}(3622.1) \rightarrow 8_1^{+}$ & $1.59(+26-20)$ & 1.491 \\
		           & $10_2^{+} \rightarrow 10_1^{+}$ & $>4.2$ & 0.065\\
		\hline ${ }^{139} \mathrm{La}$ & $5 / 2_1^{+} \rightarrow 7 / 2_1^{+} $ & $<2$ & 0.007\\
        		  & $\left(5 / 2_2\right)^{+} \rightarrow 7 / 2_1^{+}$ & 3.7(13) & 12.956 \\
		          & $9 / 2_1^{+} \rightarrow 7 / 2_1^{+}$ & 7.4(11) & $2.31 \times 10^{-5}$ \\
                   & $9 / 2_1^{+} \rightarrow 5 / 2_1^{+} $ & 1.76(24) & 7.679  \\
                   & $9 / 2_2^{+}\rightarrow7 / 2_1^{+}$ & - & 7.382 \\
                   & $9 / 2_2^{+}\rightarrow5 / 2_1^{+}$ & - & 0.055 \\
		\hline ${ }^{140} \mathrm{Ce}$ & $2_1^{+} \rightarrow 0_1^{+}$ & 13.9(7) &12.172 \\
		           & $4_1^{+} \rightarrow 2_1^{+}$ & 0.1370(12) &0.166 \\
		           & $6_1^{+} \rightarrow 4_1^{+}$ & $0.29(+8-5)$ & 0.094 \\
                        & $8_1^{+} \rightarrow 6_1^{+}$ & - 			& 		3.109\\
		           & $10_1^{+} \rightarrow 8_1^{+}$ & $0.55(14)$ & 0.345 \\
		\hline ${ }^{141} \mathrm{Pr}$ & $7 / 2_1^{+} \rightarrow 5 / 2_1^{+}$ & $0.35(+8-7)$ & 0.079\\
		           & $3 / 2_1^{+} \rightarrow 7 / 2_1^{+}$ & $<2.12$ &0.571 \\
		           & $3 / 2_1^{+} \rightarrow 5 / 2_1^{+}$ & $<8.12$ & 17.585 \\
		           & $9 / 2_1^{+} \rightarrow 5 / 2_1^{+}$ & $1.95(+47-42)$ & 0.367\\
                      & $9 / 2_2^{+} \rightarrow 5 / 2_1^{+}$ & $9.3(11)$ & 8.173\\
		           & $11 / 2_1^{+} \rightarrow 7 / 2_1^{+}$ & $6.3(+31-26)$ &7.943 \\
		           & $13 / 2_1^{+} \rightarrow 11 / 2_1^{+}$ & $<82.72$ & 0.373\\
		           & $13 / 2_1^{+} \rightarrow 9 / 2_1^{+}$ & $<338$ & 0.708 \\
		           & $15 / 2_1^{+} \rightarrow 11 / 2_1^{+}$ & $0.89(+11-9)$ & 1.117 \\
		\hline ${ }^{142} \mathrm{Nd}$ & $2_1^{+} \rightarrow 0_1^{+}$ & $12.03(22)$ & 12.972 \\
                   & $4_1^{+} \rightarrow 2_1^{+}$ & - 	& 	0.019	\\
                     & $6_1^{+} \rightarrow 4_1^{+}$ & -	& 	0.033	\\
                     & $8_1^{+} \rightarrow 6_1^{+}$ & - 	& 		4.372\\
                     & $10_1^{+} \rightarrow 8_1^{+}$ & - & 		0.588\\
		\hline ${ }^{143} \mathrm{Pm}$ & $7 / 2_1^{+} \rightarrow 5 / 2_1^{+}$ & $<0.17$ & 0.298 \\
		           & $15 / 2_1^{+} \rightarrow 11 / 2_1^{+}$ & 1.52(6) & 0.430\\
		\hline ${ }^{144} \mathrm{Sm}$ & $2_1^{+} \rightarrow 0_1^{+}$ & 11.9(4) & 12.423 \\
		          & $4_1^{+} \rightarrow 2_1^{+}$ & $<790$ & 0.600 \\
		           & $6_1^{+} \rightarrow 4_1^{+}$ & 0.188(6) & 0.106 \\
                       & $8_1^{+} \rightarrow 6_1^{+}$ & - 	&1.504\\
                     & $10_1^{+} \rightarrow 8_1^{+}$ & - & 	0.002	\\
		           & $9_1^{-} \rightarrow 7_1^{-}$ & 0.35(15) & 0.122\\
        \hline ${ }^{145} \mathrm{Eu}$ & $7 / 2_1^{+} \rightarrow 5 / 2_1^{+}$ & - & 0.617 \\
		\hline ${ }^{146} \mathrm{Gd}$ & $2_1^{+} \rightarrow 0_1^{+}$ & $>0.59$ & 10.980\\
                   & $4_1^{+} \rightarrow 2_1^{+}$ & - 	& 	2.821	\\
                     & $6_1^{+} \rightarrow 4_1^{+}$ & -	& 0.194\\
                     & $8_1^{+} \rightarrow 6_1^{+}$ & - 	& 	 0.093	\\
                     & $10_1^{+} \rightarrow 8_1^{+}$ & - & 	 0.894	\\
		          & $7_1^{-} \rightarrow 5_1^{-}$ & 0.52(7) & 0.523 \\
                \hline ${ }^{147} \mathrm{Tb}$ & $3 / 2_1^{+} \rightarrow 1 / 2_1^{+}$ & - & 3.913 \\
                & $5 / 2_1^{+} \rightarrow 1 / 2_1^{+}$ & - & 0.069 \\
		\hline ${ }^{148} \mathrm{Dy}$ & $2_1^{+} \rightarrow 0_1^{+}$ & -&14.142 \\
                       & $4_1^{+} \rightarrow 2_1^{+}$ & - 	&  5.332		\\
                     & $6_1^{+} \rightarrow 4_1^{+}$ & -	& 	 4.739	\\
                    & $8_1^{+} \rightarrow 6_1^{+}$ & 3.8(13)	& 	3.332 	\\
		           & $10_1^{+} \rightarrow 8_1^{+}$ & 0.93(7) &1.297 \\
		\hline ${ }^{149} \mathrm{Ho}$ & $\left(27 / 2_1^{-}\right) \rightarrow\left(23 / 2_1^{-}\right)$ & $1.87(+15-12)$ & 3.160\\
		\hline ${ }^{150} \mathrm{Er}$ & $2_1^{+} \rightarrow 0_1^{+}$ &  -& 15.645\\
                  & $4_1^{+} \rightarrow 2_1^{+}$ & - 	& 	1.530	\\
                     & $6_1^{+} \rightarrow 4_1^{+}$ & -	& 	1.784	\\
                    & $8_1^{+} \rightarrow 6_1^{+}$ & - 	& 	 0.495	\\
                     & $10_1^{+} \rightarrow 8_1^{+}$ & 0.24(3) & 	0.328	\\
		\hline ${ }^{151} \mathrm{Tm}$ & $\left(27 / 2_1^{-}\right) \rightarrow\left(23 / 2_1^{-}\right)$ & 0.257(14) &0.171 \\
		\hline ${ }^{152} \mathrm{Yb}$ & $2_1^{+} \rightarrow 0_1^{+}$ & - &16.043 \\
                        & $4_1^{+} \rightarrow 2_1^{+}$ & - 	& 	 0.061	\\
                & $6_1^{+} \rightarrow 4_1^{+}$ & -	& 	0.026	\\
                & $8_1^{+} \rightarrow 6_1^{+}$ & - 	& 	 0.011	\\
                & $(10_1^{+}) \rightarrow (8_1)^{+}$ & 0.020(3) & 	0.004	\\
		\hline ${ }^{153} \mathrm{Lu}$ & $\left(23 / 2_1^{-}\right) \rightarrow\left(19 / 2_1^{-}\right)$ & $<0.01$ &0.016 \\
		           & $27 / 2_1^{-} \rightarrow 23 / 2_1^{-}$ & $0.0092(+23-16)$ & 0.016\\
		\hline ${ }^{154} \mathrm{Hf}$ & $2_1^{+} \rightarrow 0_1^{+}$ & - &14.791\\
                  & $4_1^{+} \rightarrow 2_1^{+}$ & - 	& 	0.568	\\
                & $6_1^{+} \rightarrow 4_1^{+}$ & -	& 	 1.262	\\
                & $8_1^{+} \rightarrow 6_1^{+}$ & - 	& 	0.177	\\
                & $10_1^{+} \rightarrow 8_1^{+}$ & - & 		0.227\\
            \hline ${ }^{155} \mathrm{Ta}$ & $1/2_1^{+} \rightarrow 3/2_1^{+}$ & - &6.329\\
                  & $27 / 2_1^{-} \rightarrow 23 / 2_1^{-}$ & - & 2.373\\      
            \hline ${ }^{156} \mathrm{W}$ & $2_1^{+} \rightarrow 0_1^{+}$ & - &11.403\\
                  & $4_1^{+} \rightarrow 2_1^{+}$ & - 	& 	4.782\\
                & $6_1^{+} \rightarrow 4_1^{+}$ & -	& 	 4.748	\\
                & $8_1^{+} \rightarrow 6_1^{+}$ & - 	& 	2.878\\
                & $10_1^{+} \rightarrow 8_1^{+}$ & - & 		1.197\\
                         \hline ${ }^{157} \mathrm{Re}$ & $1/2_1^{+} \rightarrow 3/2_1^{+}$ & - &6.772\\
                 & $27 / 2_1^{-} \rightarrow 23 / 2_1^{-}$ & - & 2.748\\
            \hline ${ }^{158} \mathrm{Os}$ & $2_1^{+} \rightarrow 0_1^{+}$ & - &7.477\\
                  & $4_1^{+} \rightarrow 2_1^{+}$ & - 	& 	6.729	\\
                & $6_1^{+} \rightarrow 4_1^{+}$ & -	& 	 0.013\\
                & $8_1^{+} \rightarrow 6_1^{+}$ & - 	& 	4.663	\\
                & $10_1^{+} \rightarrow 8_1^{+}$ & - & 		1.697\\
                         \hline ${ }^{159} \mathrm{Ir}$ & $1/2_1^{+} \rightarrow 3/2_1^{+}$ & - &5.474\\
                 & $27 / 2_1^{-} \rightarrow 23 / 2_1^{-}$ & - & 6.251\\
		\hline\hline
\end{longtable}

\begin{longtable}{c|c|c|c}  \label{tablegfactor}  \\
\caption{ \(g\)-factors in yrast states of $N=82$ isotones. } \\
  \hline\hline Nuclide &$I^{\pi}$ &$ g_{\rm expt.}$  & $ g_{\rm theo.}$\\
 \hline $^{134}\mathrm{Te}$ &$ 2^+$ & - & 0.691\\
          &$ 4^+$ & - & 0.684\\
           &$ 6^+$&  0.847(25)& 0.697\\
          &$ 8^+$ & - & 1.265\\
         &$10^+$ & - & 1.265\\
 \hline $^{136}\mathrm{Xe}$ &$ 2^+$& 0.77(5) & 0.718\\
           &$ 4^+$& 0.80(15)& 0.695\\
           &$ 6^+$&  - & 0.694\\
           &$ 8^+$&  - & 0.694\\
           &$10^+$&  - & 0.911\\
 \hline $^{138}\mathrm{Ba}$ &$ 2^+$& 0.72(11) & 0.765\\
           &$ 4^+$& 0.80(15) & 0.712\\
           &$ 6^+$& 0.98(2) & 0.827\\
           &$ 8^+$&  - & 0.961\\
           &$10^+$& -  & 0.903\\
 \hline $^{140}\mathrm{Ce}$ &$ 2^+$& 0.95(10) & 0.992\\
          &$ 4^+$& 1.05(4) & 1.029\\
          &$ 6^+$&  - & 1.031\\
          &$ 8^+$&  - & 0.946\\
           &$10^+$& 1.03(4) & 1.018\\
          &$3^-$ &  - & 1.094\\
          &$5^-$ &  - & 1.186\\
          &$7^-$ &  - & 1.083\\
 \hline $^{142}\mathrm{Nd}$ &$ 2^+$& 0.845(75) & 1.153\\
          &$ 4^+$& -  & 1.330\\
         &$ 6^+$&  - & 1.027\\
           &$ 8^+$&   -& 1.091\\
          &$10^+$& 0.79(24) & 1.009\\
          &$3^-$ &  - & 1.089\\
          &$5^-$ &  - & 1.208\\
          &$7^-$ &  - & 1.281\\
          &$9^-$ &1.06(13)&  1.036\\
 \hline $^{144}\mathrm{Sm}$ &$ 2^+$& 0.76(11) & 1.192\\
         &$ 4^+$&  - & 1.373\\
          &$ 6^+$&  - & 1.012\\
          &$ 8^+$&  - & 0.889\\
         &$10^+$&  - & 1.239\\
         &$3^-$&  -  & 1.094\\
         &$5^-$&  -  & 1.207\\
          &$7^-$ &  - & 1.277\\
 \hline $^{146}\mathrm{Gd}$ &$ 2^+$& -  & 1.137\\
        &$ 4^+$&  - & 1.038\\
         &$ 6^+$&  - & 1.023\\
          &$ 8^+$&  - & 1.235\\
          &$10^+$&  - & 1.237\\
          &$3^-$& 0.7(3)  &1.099\\
         &$5^-$&  -  &1.198\\
         &$7^-$& 1.251(37)  &1.265\\
\hline $^{148}\mathrm{Dy}$ &$ 2^+$& -  & 1.169\\
          &$ 4^+$&  - & 1.208\\
          &$ 6^+$&  - & 1.225\\
         &$ 8^+$ &  - & 1.228\\
          &$10^+$&  - & 1.227\\
         &$3^-$&  -  & 1.096\\
         &$5^-$&  -  & 1.092\\
        &$7^-$ &  - & 1.071\\
 \hline $^{150}\mathrm{Er}$ &$ 2^+$ & - & 1.182\\
         &$ 4^+$ & - & 1.220\\
         &$ 6^+$&  - & 1.229\\
         &$ 8^+$&  - & 1.233\\
         &$10^+$&  - & 1.233\\
          &$3^-$&  -  &1.092\\
         &$5^-$ &  - &1.084\\
        &$7^-$ &   - &1.068\\
 \hline $^{152}\mathrm{Yb}$ &$ 2^+$&  - & 1.183\\
         &$ 4^+$ & - & 1.221\\
         &$ 6^+$&  - & 1.231\\
         &$ 8^+$ &  - & 1.235\\
          &$10^+$&  - & 1.236\\
         &$3^-$ &  - &1.089\\
          &$5^-$&   - &1.082\\
          &$7^-$&  -  &1.069\\
 \hline $^{154}\mathrm{Hf}$ &$ 2^+$& -  & 1.179\\
         &$ 4^+$&  - & 1.219\\
         &$ 6^+$&  - & 1.230\\
          &$ 8^+$&  - & 1.234\\
          &$10^+$&  - & 1.237\\
         &$3^-$ & - &1.088\\
         &$5^-$&   - &1.082\\
         &$7^-$ &  - &1.071\\
 \hline $^{156}\mathrm{W}$ &$ 2^+$&  - & 1.172\\
         &$ 4^+$ & - & 1.215\\
          &$ 6^+$& -  & 1.226\\
        &$ 8^+$&  - & 1.230\\
         &$10^+$& -  & 1.234\\
          &$3^-$&  -  &1.090\\
         &$5^-$ & -  &1.083\\
         &$7^-$&  -  &1.074\\
 \hline $^{158}\mathrm{Os}$ &$ 2^+$& -  & 1.032\\
           &$ 4^+$&  - & 1.086\\
          &$ 6^+$ & - & 1.231\\
           &$ 8^+$&  - & 1.240\\
           &$10^+$&  - & 1.234\\
           &$3^-$& -  &1.104\\
        &$5^-$ & -  &1.080\\
        &$7^-$ & - &1.076\\
\hline $^{133}\mathrm{Sb}$ &$ 1/2^+$ & - &  3.910\\
         &$ 3/2^+$&  - &0.418 \\
       &$ 5/2^+$ & - & 1.582\\
        &$ (7/2^+)$ &0.8571(3) & 0.677\\
         &$11/2^-$&  - &1.265 \\
 \hline $^{135}\mathrm{I}$ &$ 1/2^+$ & - & 3.032\\
         &$ 3/2^+$&  - & 0.676\\
       &$ 5/2^+$ &  -& 0.969\\
        &$ 7/2^+$ &0.8400(6) & 0.679\\
         &$11/2^-$& - & 1.259\\
 \hline $^{137}\mathrm{Cs}$ &$ 1/2^+$ & - & 2.902\\
         &$ 3/2^+$ & - & 0.654\\
           &$ 5/2^+$&  - & 1.563\\
          &$ 7/2^+$& 0.81180(3) & 0.680\\
          &$11/2^-$&  - & 1.253\\
 \hline $^{139}\mathrm{La}$ &$ 1/2^+$&  - & 2.872\\
          &$ 3/2^+$&  - & 1.025\\
           &$ 5/2^+$&  - & 1.557\\
           &$ 7/2^+$& 0.7951556(3) & 0.685\\
          &$11/2^-$&  - & 1.252\\
 \hline $^{141}\mathrm{Pr}$ &$ 1/2^+$&  - & 3.512\\
           &$ 3/2^+$&  - & 1.273\\
          &$ 5/2^+$ & 1.7102(2)& 1.537\\
           &$ 7/2^+$&  0.843(26)& 0.695\\
           &$11/2^-$& 1.13(7) & 1.249\\
           &$15/2^+$& 1.07(23) & 1.117\\
 \hline $^{143}\mathrm{Pm}$ &$ 1/2^+$ &  -& 3.392\\
           &$ 3/2^+$&  - & 1.071\\
           &$ 5/2^+$& 1.52(20) & 1.504\\
           &$ 7/2^+$&  - & 0.707\\
           &$11/2^-$& 1.236(72) & 1.247\\
           &$15/2^+$& 1.027(53) & 1.105\\
 \hline $^{145}\mathrm{Eu}$ &$ 1/2^+$&  - & 3.158\\
           &$ 3/2^+$&  - & 0.507\\
           &$ 5/2^+$& 1.597(3) & 1.467\\
           &$ 7/2^+$&  - & 0.719\\
           &$11/2^-$& 1.356(9) & 1.242\\
 \hline $^{147}\mathrm{Tb}$ &$ (1/2^+)$& 3.42(10) &2.916\\
           &$ 3/2^+$&  - &0.534\\
           &$ 5/2^+$&  - &1.481\\
          &$ 7/2^+$ & - &0.717\\
           &$11/2^-$&  - & 1.234\\
\hline $^{149}\mathrm{Ho}$  &$ 1/2^+$&  - & 3.090\\
           &$ 3/2^+$&  - & 0.521\\
         &$ 5/2^+$ &  -& 1.490\\
         &$ 7/2^+$&  - & 0.719\\
          &$11/2^-$&  -& 1.238\\
 \hline $^{151}\mathrm{Tm}$ &$ 1/2^+$&  - &3.166\\
         &$ 3/2^+$ & - &0.515\\
          &$ 5/2^+$&  - &1.494\\
         &$ 7/2^+$&  - &0.739\\
          &$11/2^-$&  - &1.240\\
 \hline $^{153}\mathrm{Lu}$ &$ 1/2^+$&  - & 3.188\\
           &$ 3/2^+$ & - & 0.515\\
          &$ 5/2^+$&  - & 1.493\\
          &$ 7/2^+$&  - & 0.839\\
         &$11/2^-$ &  -& 1.241\\
 \hline $^{155}\mathrm{Ta}$ &$ 1/2^+$&  - &3.170\\
          &$ 3/2^+$&  - &0.518\\
        &$ 5/2^+$ &  -&1.483\\
          &$ 7/2^+$&  - &0.919\\
         &$11/2^-$ &  -&1.239\\
 \hline $^{157}\mathrm{Re}$ &$ 1/2^+$ &  -&3.108\\
         &$ 3/2^+$&  - &0.526\\
         &$ 5/2^+$&  - &1.459\\
         &$ 7/2^+$&  - &0.932\\
         &$11/2^-$&  - &1.236\\
\hline $^{159}\mathrm{Ir}$  &$ 1/2^+$ &  -&3.015\\
         &$ 3/2^+$&  - &0.533\\
          &$ 5/2^+$ &  -&0.967\\
          &$ 7/2^+$&  - &0.921\\
         &$11/2^-$&  - &1.243\\
 \hline\hline
\end{longtable}

\section{The mean-field single-particle energy derived from the monopole interaction} \label{appb}

For valence protons, the monopole interaction can be written as~\cite{RevModPhys.77.427}
\begin{equation}
\hat{V}_m  =\frac{1}{2} \sum_{a b} V_m(a b) \hat{n}(a)\left[\hat{n}(b)-\delta_{a b}\right] .
\end{equation}
The contribution of the single-particle energies and the monopole interaction to the energy of an eigenstate with $n$ valence protons can be evaluated approximately as
\begin{equation}\small
\begin{aligned} 
& E[n] \equiv \sum_a \varepsilon(a) \langle\hat{n}(a)\rangle +\frac{1}{2} \sum_{a b} V_m(a b)\left\langle\hat{n}(a)\left[\hat{n}(b)-\delta_{a b}\right]\right\rangle \\
& \qquad \approx \sum_a \varepsilon(a) \langle\hat{n}(a)\rangle+\frac{1}{2} \sum_{a b} V_m(a b)\langle\hat{n}(a)\rangle\left[\langle\hat{n}(b)\rangle-\delta_{a b}\right] .
\end{aligned}
\end{equation}

In Refs.~\cite{Otsuka_2013,RevModPhys.92.015002}, the effective single-particle energy (ESPE) for orbital $a$ is defined as the energy change when one nucleon is added to that orbital in an $n$-particle system:
\begin{equation}
\varepsilon_{\mathrm{ESPE}}(a) \equiv E[n+1;a]-E[n] .
\end{equation}
Calculating the difference gives
\begin{equation}
\varepsilon_{\mathrm{ESPE}}(a)  = \varepsilon_a +  \sum_b  V_m(a b) \braket{\hat{n}(b)} .
\end{equation}
This definition corresponds to a ``one-nucleon addition energy'' and is suited for discussing observables such as one-nucleon separation energies.

In this work we adopt a similar, but distinct, quantity, the mean-field single-particle energy (MSPE). It is defined by requiring that the contribution of the single-particle energies and the monopole interaction be expressed as
\begin{equation}
E[n] = \sum_a \varepsilon_{\mathrm{MSPE}}(a)\langle\hat{n}(a)\rangle .
\end{equation}
This condition leads to
\begin{equation}
\varepsilon_{\mathrm{MSPE}}(a) = \varepsilon_a + \frac{1}{2} \sum_b  V_m(a b) \left[ \braket{\hat{n}(b)} - \delta_{ab} \right].
\end{equation}
The MSPE represents the effective spherical mean-field potential felt by each valence nucleon. It is appropriate for discussing, for example, the formation and evolution of the $Z=64$ subshell gap.

\bibliographystyle{apsrev4-2}
\bibliography{bib}

\end{document}